\documentclass{aa} % for a paper on 1 column  
\pdfoutput=1
\usepackage{graphicx}
\usepackage{epstopdf}
\usepackage{txfonts}
\newcommand{\EQ}{\begin{equation}}
\newcommand{\EN}{\end{equation}}
\newcommand{\EQA}{\begin{eqnarray}}
\newcommand{\ENA}{\end{eqnarray}}
\newcommand{\Fig}[1]{Fig.~\ref{#1}}
\newcommand{\Figu}[1]{Figure~\ref{#1}}

\newcommand{\Sec}[1]{Section~\ref{#1}}
\newcommand{\Tab}[1]{Table~\ref{#1}}
\newcommand{\BB}{{\bf {B}}}

\newcommand{\kms}{km\,s$^{-1}$}

\newcommand{\VV}{{\bf {V}}}

\def\kms{\rm km\;s$^{-1}$}

\begin{document} 

\title{Active region upflows: 2. \\
       Data driven MHD modeling}

\author{K. Galsgaard\inst{1}
        \and
        M. S. Madjarska\inst{2}
        \and
        K. Vanninathan\inst{3}
        \and 
        Z. Huang\inst{4}
        \and
        M. Presmann\inst{1}
}

\institute{Niels Bohr Institute\\
           Geological Museum\\
           {\O}stervoldgade 5-7, 1350 Copenhagen K, Denmark\\
           http://www.researcherid.com/rid/A-7793-2012 \\
           \email{kg@nbi.ku.dk}
           \and
           Armagh Observatory, College Hill, Armagh BT61 9DG, N. Ireland\\
           \email{madj@arm.ac.uk} 
           \and
           Institute of Physics/IGAM, University of Graz, 8010 Graz, Austria\\
           \email{kamalam.vanninathan@uni-graz.at}
           \and
           Shandong Provincial Key Laboratory of Optical Astronomy and Solar-Terrestrial
           Environment, Institute of Space Sciences, Shandong University, Weihai, 264209
           Shandong, China \\  
           \email{huangzhenghua@gmail.com}
}

\date{\today}

\abstract 
   {
    Observations of many active regions show a slow systematic outflow/upflow
    from their edges lasting from hours to days. At present no physical explanation 
    has been proven, while several suggestions have been put forward. 
   }
   {
    This paper investigates one possible method for maintaining these upflows 
    assuming that convective motions drive the magnetic field to initiate 
    them through magnetic reconnection.
   }
   {
    We use Helioseismic and Magnetic Imager (HMI) data to provide an initial 
    potential three dimensional magnetic field of the 
    active region NOAA 11123 on 2010 November 13 where the characteristic 
    upflow velocities are observed. A simple one-dimensional hydrostatic 
    atmospheric model covering the region from the photosphere to the corona
    is derived. Local Correlation Tracking of the magnetic features
    in the HMI data is used to derive a proxy for the time dependent velocity
    field. The time dependent evolution of the system is solved using a resistive 
    three-dimensional MagnetoHydro-Dynamic code.
   }
   {
    The magnetic field contains several null points located well above the 
    photosphere, with their fan planes dividing the magnetic field into  
    independent open and closed flux domains.
    The stressing of the interfaces between the different flux
    domains is expected to provide locations where magnetic reconnection can
    take place and drive systematic flows. In this case, the region between
    the closed and open flux is identified as the region where observations
    find the systematic upflows.
   }
   {In the present experiment, the driving only initiates magneto-acoustic 
    waves, without driving any systematic upflows at any of the flux interfaces.
    }

%%%% check the official list !!!
% http://www.aanda.org/index2.php?option=com_content&task=view&id=170&Itemid=256

\maketitle
%
%________________________________________________________________

\section{Introduction}

The Extreme ultraviolet Imaging Spectrometer (EIS) on-board Hinode provides detailed observations of active regions, showing the existence of localised regions with systematic, relative low velocity upflows that last from hours to days. However, neither the source region nor the mechanism which drives these upflows have been identified yet although several theories have been put forward. 
There has been a general consensus to use the expression "outflows" for the phenomenon, with the indirect assumption that these flows are related to open field lines that allow the upflows to contribute directly to the slow solar wind. As most models have been based on local field extrapolations, it is not certain these field lines are really open and allow the plasma to escape the sun. We, therefore, refer to these flows as {\it upflows} in this paper.
For a detailed review of the observational contributions to this field see \citet[][hereafter Paper I]{Vanninathan_ea_15}.

% ??? incl information on the obs paper here or in the discusssion ???

Over time different theories have been suggested to explain the mechanisms that drive the upflows. It has been suggested that 3 and 5 minutes oscillations can drive upflows \citep{DelZanna2008b, Marsch2004}. This type of oscillations have been identified in the upflow regions by \citet{2010RAA....10.1307G} and \citet{Ugarte-Urra2011}, who propose the underlying processes to be sporadic in both space and time. 
From a potential magnetic field extrapolation, \citet{Harra2008} found that reconnection on one side of the investigated active region generated long loops which connect across the region. They concluded that the observed upflows were the result of an expansion of the newly formed loops as they move up, towards a new equilibrium. 

In a different investigation \citet{2008ApJ...685.1262M} identified circulation of plasma through magnetic funnels and closed loops as a natural state of the plasma that continuously is in motion. They suggested the photospheric convection as the ultimate driving mechanism of the plasma motions in the corona. Using potential field extrapolations of a region containing two active regions \citet{Boutry2012} found that about 18\% of the mass upflow from one region was accounted for as downflow in the neighbouring region. 
\citet{2011A&A...526A.137D} found a relation between radio noise storms and coronal upflows, indicating that magnetic reconnection may be initiating the process. From magnetic field extrapolations they find magnetic null points and associated separatrices in the corona which could be associated with magnetic reconnection. A study of an active region-coronal hole (AR-CH) boundary by \citet{vanDriel-Gesztelyi2012} supports the hypotheses that upflows are caused by magnetic reconnection along null separatrices. They identify a magnetic null point in the corona and propose this to be the site of continuous reconnection. 
\citet{DelZanna2008b} speculates that the observed blue-shifts could be the result of chromospheric evaporation driven by magnetic reconnection. The author hypothesizes that this mechanism is similar to the one observed during the gradual phase of two ribbon flares. Here, the upflows are seen from the footpoints of the newly formed loops as a result of the indirect heating from the reconnection point high above the photosphere. 
\citet{2014SoPh..289.4501S} made a simple MHD simulation of a stratified loop and imposed an excess pressure at the photospheric region. Their experiment shows that upflow velocities within the observed range can be generated by this mechanism.
% That an excess pressure can cause an upflow within the observed velocity range was shown by \citet{2014SoPh..289.4501S} who made a simple MHD simulation of a stratified loop and imposed an excess pressure at the photospheric region of the loop.
\citet{Baker2009} combined EIS observations and constant alpha force free magnetic field extrapolation based on Helioseismic and Magnetic Imager (HMI) data to study the magnetic field structures relation to the upflow regions. They identify Quasi Separatrix Layers (QSLs) \citep{1995JGR...10023443P} between short closed and longer "open" field lines to be located in the vicinity of the upflow region and propose reconnection to take place along or near the QSLs. This magnetic field evolution can result in smooth extended upflow region as long as the QSLs are continuously stressed. Depending on the driver and field structure this model could generate a host of different effects such as chromospheric evaporation, reconnection jets, siphon flows along closed loops and upflows due to pressure gradients generated after the reconnection between the two different loops. 

By studying an AR-CH complex through numerical simulations \citet{2010SoPh..261..253M} was able to model the upflows as a result of a new magnetic flux emergence. The new flux will push its way into the coronal hole region where it initiates compression waves on the open field lines. This short duration event is a precursor for the build up of a local current sheet where reconnection will be initiated. 
%The downside is that this phase only last as long as the emerging region is expanding into the coronal material and a reconnection process has not been initiated. It can therefore not explain the long duration events seen in observations.
Following this \citet{Harra2010} found new flux emergence to be associated with a decaying active region that displayed upflows. \citet{Harra2012} investigated the role of a new emerging flux in producing upflows through MHD simulations where a new bipole emerges into an open magnetic field configuration. They conclude that the observed blue-shifts are either reconnection jets or pressure-driven upflows. The two MHD models \citep{2010SoPh..261..253M,Harra2012} used in these investigations are identical and only the extent of the interpretation of the results differ slightly.

Although several indirect observations \citep{McIntosh2009,He2010,Nishizuka2011} have indicated a correlation between jets in the chromosphere and coronal upflows, in Paper I it has been shown through chromospheric spectroscopic imaging that the upflows persists despite the absence of chromospheric jets, indicating that the driving should be located higher in the atmosphere.

%CULHANE 2014
There has been a general assumption that the upflow takes place along open magnetic field lines, allowing the plasma to directly contribute to the slow solar wind. An analysis of an active region by \citet{2014SoPh..289.3799C} shows this is not a necessary case. Their global potential field source surface (PFSS) model (which is the only investigation taking into account the large scale magnetic field) suggests that a QSL related reconnection takes place close to the active region populating the field lines with active region FIP enhanced plasma. These field lines are further required to pass through a second reconnection with a distant 3D null, experiencing a spine-fan reconnection \citep{2009PhPl...16l2101P, 2013ApJ...774..154P, 2011A&A...529A..20G} process before the plasma finally enters an open field region.
This requires a very precise line up of the different field line systems that seems unlikely for the typical upflow regions and even a change from a potential to a non-linear force free field (NLFFF) model would most likely remove this alignment.

In this paper we investigate how restricted photospheric convective motions (only horizontal flow components) may  assist in creating a systematic upflow in an active region  by driving a localised reconnection process. A data driven MHD model is used in the investigation, where a time sequence of magnetograms are used to provide an assumed boundary driving flow, and the initial magnetic field structure is provided by a potential field extrapolation from a HMI magnetogram. The layout of the paper is as follows. In \Sec{init.sec} the initial conditions and the numerical approach are discussed. The results are discussed in \Sec{res.sec}. \Sec{dis.sec} presents the results relative to the observations. The conclusions are given in \Sec{con.sec}.
%________________________________________________________________
\section{Initial conditions}
\label{init.sec}

% HMI data and driving flows
%% fill in much more information her, both about the specific case.
The data driven experiment presented in this paper is based on the observational study of NOAA 11123 analysed in Paper I. This investigation shows the southern edge of the active region to host a systematic upflow with a characteristic speed of 5 -- 20~\kms\ but also a second and much faster component of 105~\kms.
The upflow lasts for at least the 3.5 hours period the observations cover and it does not seem to be initiated by magnetic flux emergence, but more likely by some effect that acts well above the chromosphere, but at the same time allows for ample diffusion of photospheric magnetic field (see Paper I). Having detailed observations by the Solar Dynamical Observatory (SDO) Atmospheric Imaging Assembly (AIA) and HMI makes the event an ideal target for a data driven investigation, aiming at understanding the physical implications of the photospheric driving on the systematic upflow.

\begin{figure}
   \centering
   \includegraphics[width=7cm]{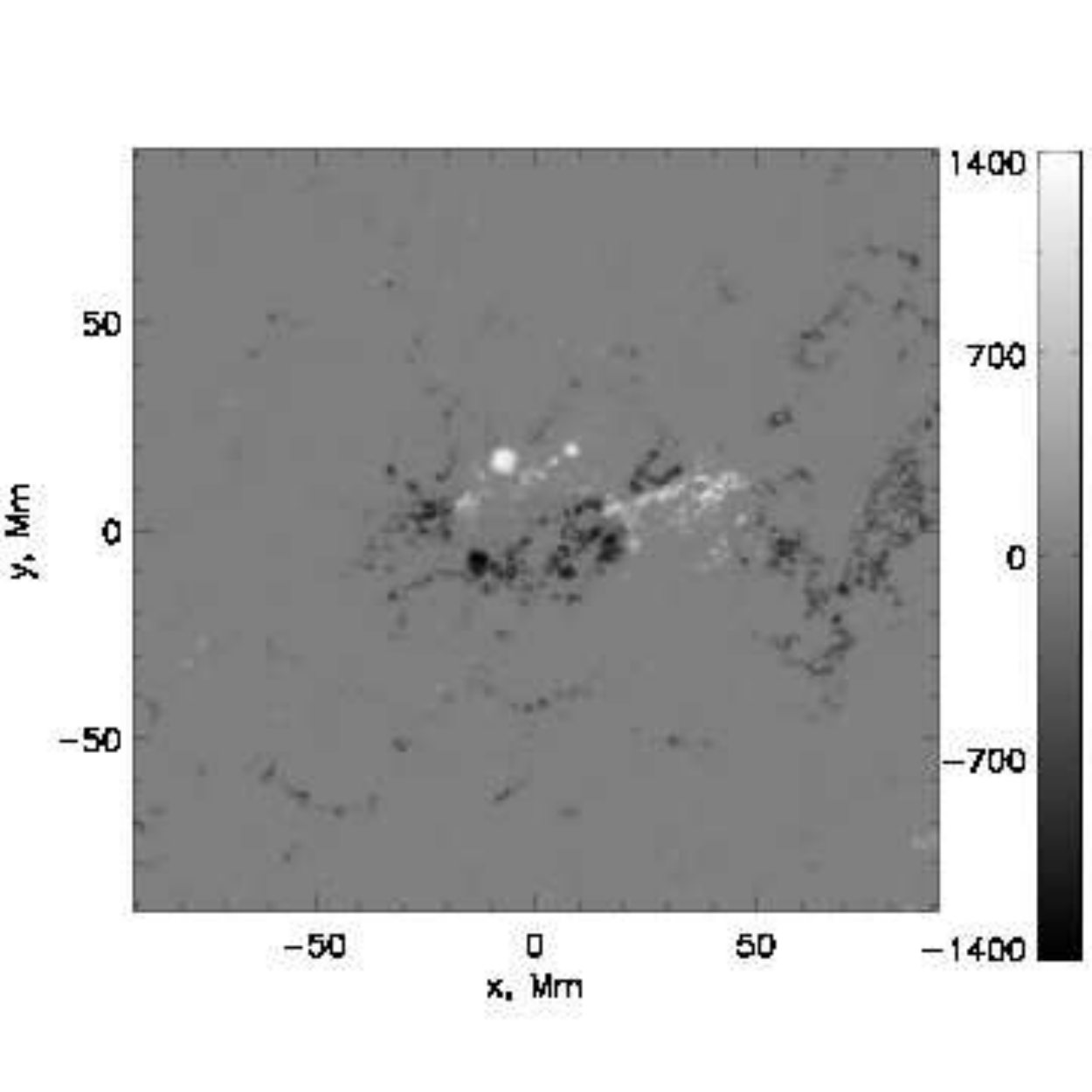} \,

   \includegraphics[height=5.2cm]{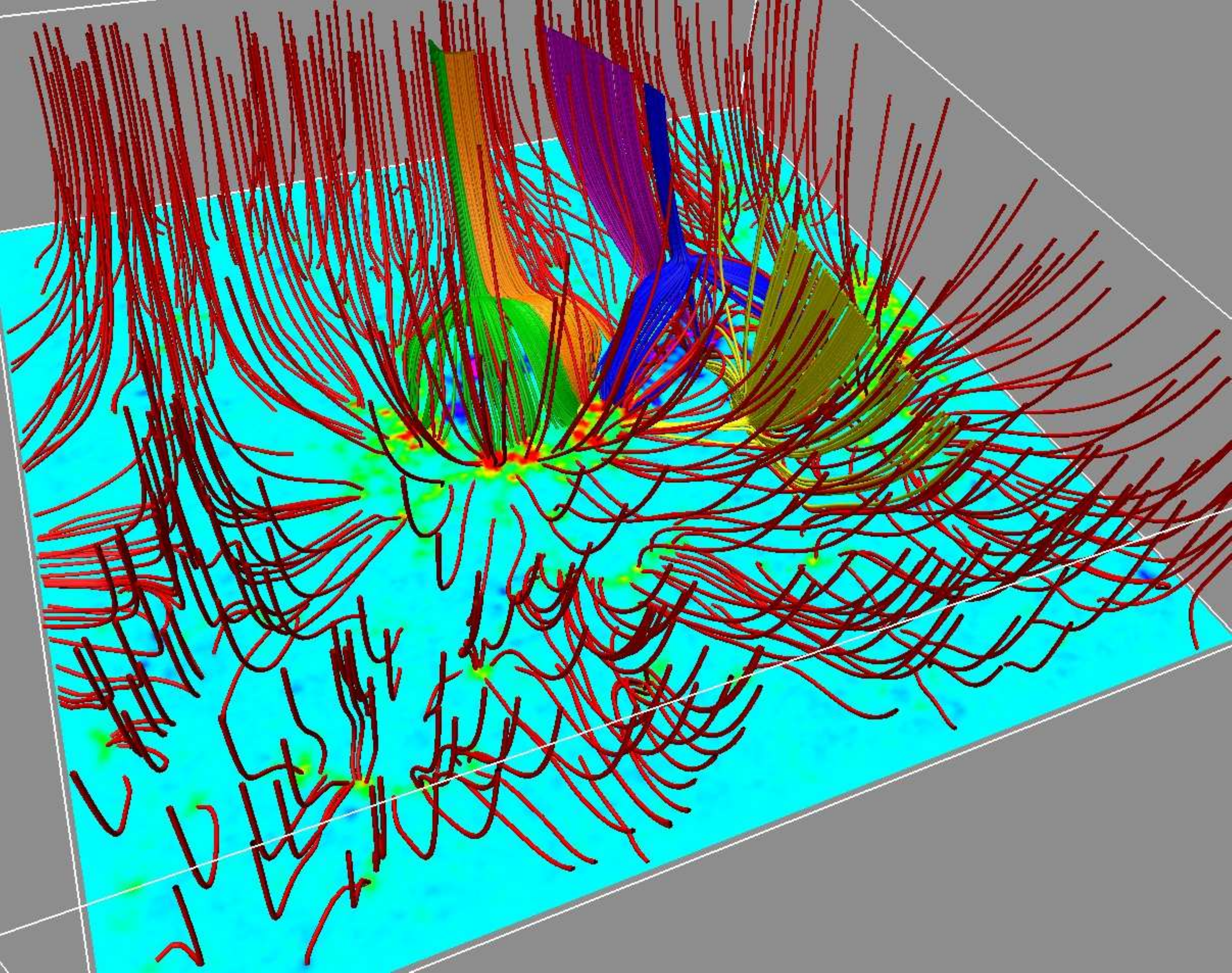}
   \caption{TOP: Initial HMI magnetogram used to derive the potential magnetic field. 
            The dynamic range is $\pm$ 1400 G.
            BOTTOM: 3D field line structure of the closed region areas with 
            field lines traces from the vicinity of the dominating null points (green, orange,
            red, blue and yellow).
            These are combined with field lines of the "open" field (red) region located around 
            the active region. The orientation is rotated slightly relative to the 
            magnetogram, with the left, right and up down edges being the same in the
            two representations. The size of the magnetogram is 256"$\times$256".   
           }
   \label{fig1.fig}
\end{figure}

To preform a data driven 3D MHD model, several independent steps are required. The line of sight HMI observations provide a time series of magnetograms with a time resolution of 45 s. These data allow for either a potential or a constant alpha force free 3D model extrapolations of the magnetic field structure. From the available full disk HMI data a suitable region around the active region was selected to limit edge effects that may influence the extrapolated magnetic field structure. A NLFFF extrapolation was also attempted, but due to the location of the active region and its strong flux imbalance, the result was unsatisfactory for further analysis (private discussion with Thomas Wiegelmann). The selected region is shown in the top panel of \Fig{fig1.fig}, where the line of sight magnetic field component at the edges of the region is smoothed to zero over a short distance. This allows for a horizontally 2D periodic three-dimensional (3D) magnetic field extrapolation. The HMI magnetograms are de-rotated to the same physical time, making a direct comparison between the time dependent structure of the region easier (see Paper I for more details).

Potential magnetic field extrapolations are obtained using a Fast Fourier Transform (FFT) approach. The first of these field extrapolations was used as the initial magnetic field for the MHD experiment. Before it can be used in the MHD code, it has to be further manipulated, making sure it is potential and divergence free when applying the high order stencil adopted in the code. A representation of the 3D magnetic field structure is shown in the bottom panel of \Fig{fig1.fig}. This shows selected field lines, traced both from the vicinity of the most important 3D null points in the domain, and from randomly scattered points on the photospheric surface. 

A full discussion of the field topology is provided below. Combining model field-line traces with the information about the location of the upflow region identified in Paper I, shows that the upflow region originates from the interface region between the open and closed flux region bounded by the green, orange and blue field lines (see \Fig{null_field.fig} below.). Here, open field lines are only to be taken as indicative as the extrapolation is conducted over a smaller region of the Sun not taking into account active regions further away. 
%fig 1

%HMI data are available for this region covering 3.5 hours throughout which the outflow is constantly observed \citep{Vanninathan_ea_15}. 
A potential field model is not the most appropriate starting condition for the simulation as there is no free magnetic energy and this has to be build up before the magnetic field can start driving the upflow. When comparing the extrapolated field line structure with EUV observations (AIA 171~{\AA} and 193~{\AA}), only minor variations between the two extrapolations are seen, and redoing the model calculations with a constant alpha force free magnetic field does not increase significantly the amount of free magnetic energy. This also has the problem that this type of extrapolation requires flux balance on the photospheric boundary to be performed which in the present example would significantly change the magnetic field configuration.
% The reason for this is that the main fitting is done on the large scale structures of the active region, while the most significant differences between a potential and a constant alpha field are found on the smaller scales. 

\begin{figure}
\begin{center}
\includegraphics[width=8cm]{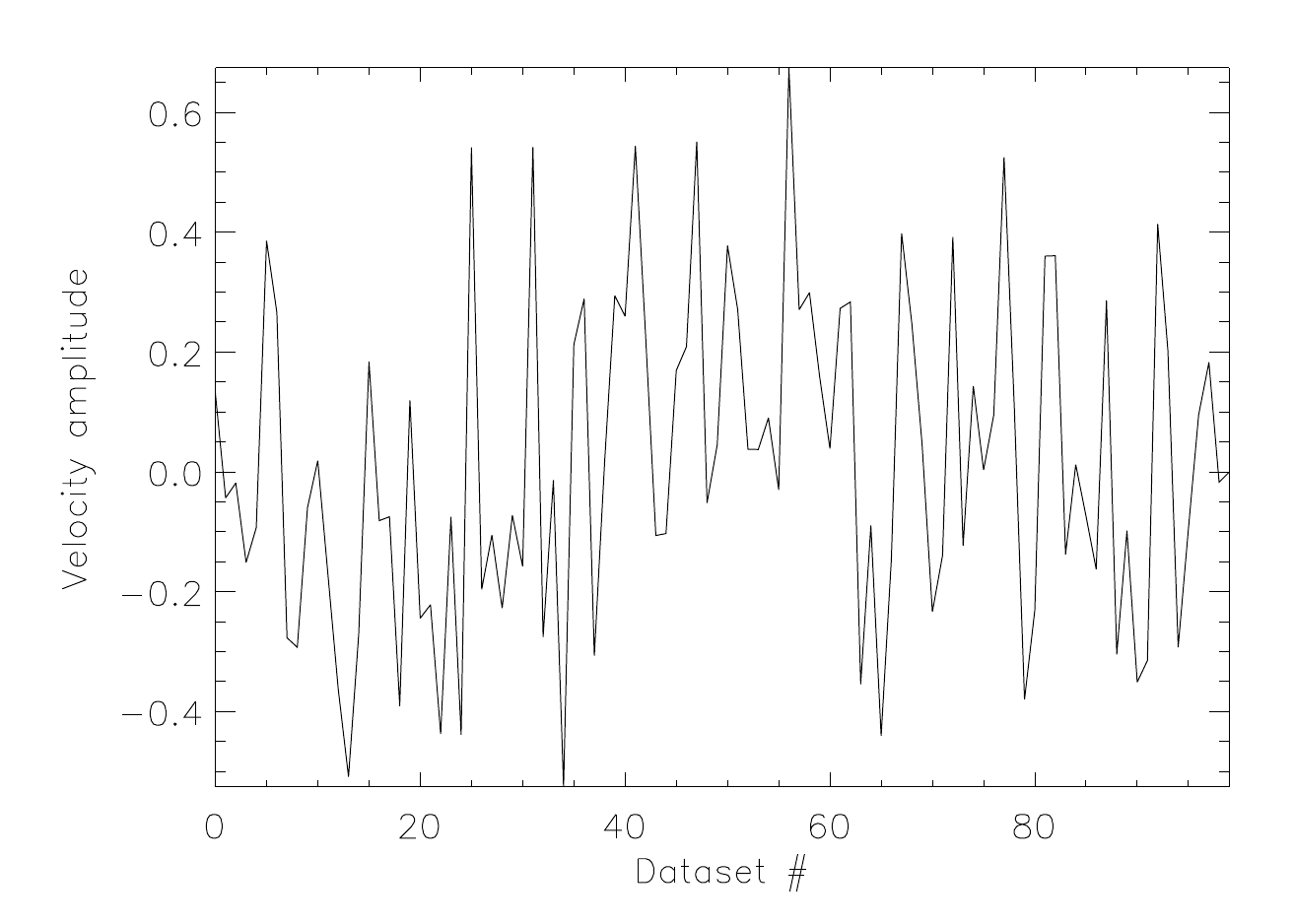}
\end{center}
\caption[]{\label{pixel_velocity_time.fig} The change of one velocity component 
           (in arbitrary units) as a function of HMI snapshot number (time) for a 
           randomly chosen pixel in the plane.
           }
\end{figure}

To stress the magnetic field configuration, a boundary flow is derived using the {\it Local Correlation Tracking} (LCT) algorithm developed by \citet{2008ASPC..383..373F}. This algorithm utilises the information provided by the changes in the positions of the magnetic fragments between two consecutive HMI magnetograms to obtain representative horizontal velocity vectors for that time interval. Repeating this on the whole dataset provides a time dependent 2D horizontal velocity flow that to a given accuracy represents the change in the magnetic field positions as a function of time.
In applying the LCT approach one needs to choose a stencil width over which the correlation is done. This value is critical for two reasons. First, its size determines the physical size of the flow structures that may be determined by the algorithm. Second, choosing a very small stencil size allows small scale flow structures to be detected, while at the same time the algorithm becomes critically dependent on noise in the data. This can provide large fluctuations in the flow field within very short distances. This high amplitude noise subsequently has to be smoothed/filtered out before using the data for driving the model. 
%Examples of a velocity component can be seen in \Fig{driver_sigma.fig}, showing the impact of the magnitude of the sigma parameter.
%\begin{figure}
%\begin{center}
%\includegraphics[width=6cm]{eps/driver_19.eps}
%\includegraphics[width=6cm]{eps/driver_15.eps}
%\includegraphics[width=6cm]{eps/driver_10.eps}
%\includegraphics[width=6cm]{eps/driver_05.eps}
%\end{center}
%\caption[]{\label{driver_sigma.fig} Four examples of the characteristic flow profiles obtained by varying the LCT codes sigma parameter. These represents the $v_x$ component. From top left to bottom right the applied sigma values are 19, 15, 10 and 5 gridpoints. The same absolute scaling is used for each of the frames to show the changing structure of the velocity flow. The white/black indicates a positive/negative velocity component. The domain represents the same area as show in the left frame of \Fig{fig1.fig}.
%}
%\end{figure}
The increasing noise level with a decreasing sigma value puts a limit on how small length scales the velocity structures determined by this method may have. 
Comparing with the underlying HMI data, it is clear that a lot of small scale motions exist in these data. These motions are responsible for the rapid advection of small fragments, representing typically a diffusion-like processes of the magnetic field structure. In the present approach we, naively, assume small scale motions are unimportant for the general evolution of the magnetic field, and therefore we apply the velocity field determined by the standard sigma=19 value.
 These small scale motions naturally have implication on introducing small scale "noise" on top of the large scale advection of the magnetic footpoints. If they are important for stressing the global system, we are missing this energy input in this simulation.
We use the full cadence of the HMI data, i.e. 45 s. The small scale convective motion does not change dramatically on this time scale (45 s) as the typical lifetime of the granulation is 5 min, i.e. around 6 times longer. One would, therefore, expect the LCT applied on a full cadence HMI dataset to provide a velocity profile that slowly changes with time. The length scale that is resolved with the applied sigma value, is much larger than the granular length scale and does represent an even longer time scale in the data. This makes it surprising to find a velocity profile that changes dramatically between each data set and that this result seems to be independent on the choice of sigma smoothing of the LCT algorithm. An example of the time dependent changes in a single velocity component for a single pixel is shown in \Fig{pixel_velocity_time.fig}.  This may be the result of the noise in the independent datasets that gives rise to problematic velocity information. The critical is that we still find similar fluctuations in the flow determination when a longer time base between observations is used in deriving a velocity profile.

Before the driving profile is used in the experiment, a simple smoothing of the data has been done over time to limit the largest and most rapid changes in the flow profile. Further to this, the flow velocity, like the magnetic field, is damped towards zero at the physical boundaries of the domain to allow it to be consistently used in a 2D periodic domain. These velocity data are used as input for the imposed boundary driving for the MHD experiment described below. The driver information is defined at discreet time intervals of 45 s, a time interval that is much larger than the time stepping used in the MHD code. To provide a smooth change in the driving velocity as a function of time a third order interpolation algorithm is applied to the time varying LCT velocity to get the actual flow vector at a given time.

\begin{figure}
   \centering
   \includegraphics[width=8cm]{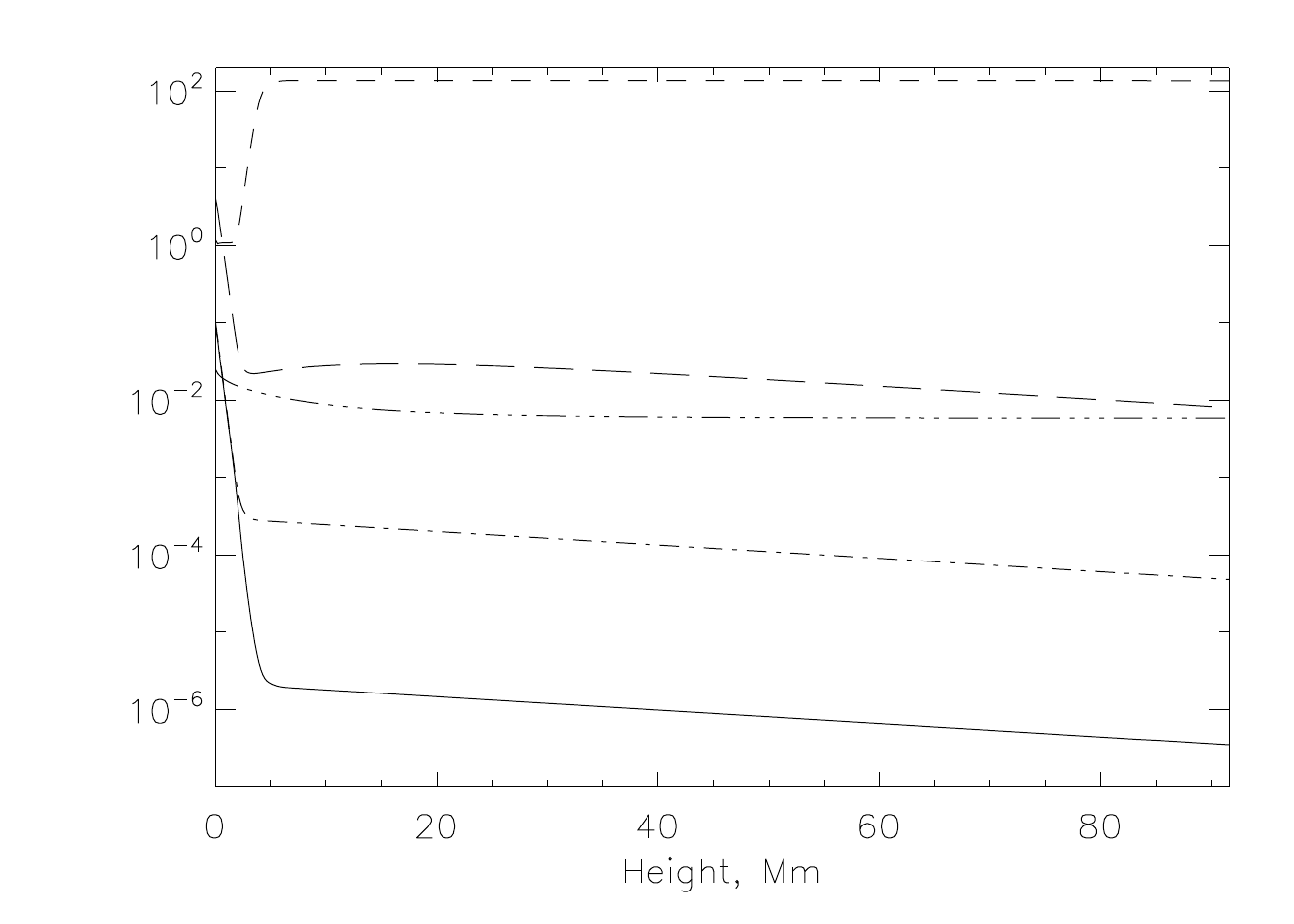}
   \caption{ 1D hydrostatic model atmosphere parameters in code units as functions of
             height (Mm), showing the density (solid line), the temperature 
             (dashed line) and the gas pressure (dot-dashed line). For
             comparison, the average magnetic field strength is shown as the 
             triple-dotted-dashed line with the long-dashed line representing
             the average plasma beta of the model.
           }
   \label{fig2.fig}
\end{figure}

% Hydrostadic equlibrium
As a supplement to the potential magnetic field a simple hydrostatic 1D model of the solar atmosphere is applied. The domain covers the region from the photosphere and into the corona. The model is defined by specifying the temperature profile with height and then integrating the hydrostatic equation using a simple integration algorithm. Subsequently, this profile is iterated to fit the hydrostatic criteria using the chosen grid spacing and the high order finite difference algorithm applied in the MHD code. For simplicity the plasma is assumed to follow the ideal gas law, and optical thin radiation and heat conduction are ignored in the energy equation. 
These terms are ignored in the present investigation, as we are  testing the idea that a magnetic reconnection process directly provide the reason for the continuous upflow. The main aim is, therefore, to look for these processes that to a very large extent develop independently of the accuracy of the energy equation. When starting from a potential magnetic field and a simple 1D hydrostatic equilibrium, one needs the magnetic field to be stressed for a period of time before it reaches a level where it is possible to ignore an artificial background heating source to maintain a hot corona. From the analysis in Paper~I it is clear that the upflow does not originate from below the lower transition region, where most evaporation flows are initiated. Including these terms in the investigation is therefore a direct contradiction to what we are trying to do, namely a direct data simulation where the  magnetic field is to be stressed in a way that to a feasible degree represents the driving in the observations and hopefully ends driving a reconnection process.
\Fig{fig2.fig} shows a 1D representation of the field quantities imposed on the initial conditions. 
 \cite{2014PASJ...66S...7B} did a similar experiment with a different active region. There they show that to include heat conduction and radiation from this type of initial conditions takes on the order of an 1 hour solar time to reach an energetic steady state. After this the calculations become expensive due to the restriction on the time step from the heat conduction. How this influences the magnetic structure and its ability to dissipate magnetic energy through reconnection has not been investigated.

%fig 2

% MHD approach
The experiment is evolved in time by solving the non-ideal MHD equations (as given in the paper by \citet{2013SoPh..284..467B}) using the Copenhagen Stagger Code which contains hyper-diffusive viscous and ohmic terms self-consistent in the momentum, energy and induction equations \citep{Nordlund_Galsgaard1997}. The numerical box has $512 \times 512 \times 511$ points in the $(x,y,z)$ directions defined on a uniform grid. The model represents a domain of size $184.3 \times 182.3 \times 91.4$ Mm. This gives a grid resolution of 0.36 Mm in the horizontal direction and 0.18 Mm in the vertical direction. 
The scaling between the code units and real units are given by setting $t_0=100$ sec, $L_0=10^6$ m and $\rho_0=10^{-4}$ kg/m$^3$ from which it follows that $T_0=10^4$ K and $B_0=1121$ Gauss. 
%
% Boundary conditions
The boundary conditions of the domain are periodic in the two horizontal directions ($x,y$) for simplicity. The domain is closed in the vertical direction ($z$). This prevents mass flow in/out of the domain and limits the driving velocity on the lower boundary to be only the horizontal flows defined by the LCT approach. As the LCT approach provide a proxy for the flow speed and structure, we take the freedom in the experiment, namely to allow changes in the imposed driving amplitude, and the time interval between the different datasets. Such changes allow for the investigation of different characteristics in the flow patterns impact on the dynamical evolution of the magnetic field configuration and is a typical way to speedup the slow driving time relative the coronal transit time. This is often used to save computing time and to prevent diffusion releasing the imposed free energy before it has reached a level where more dynamical phenomena can take place. This is a common issue when conducting numerical experiments where the diffusion is many times larger than in the real solar environment. \Tab{t1.tab} contains the driver parameters used for the various models discussed in the paper. Additional to the zero normal velocity component imposed on the top and bottom boundaries, a dedicated damping layer is implemented in a smaller part close to the top of the domain. This prevents upward propagating waves to be reflected downwards by the imposed zero normal boundary velocity. The damping is implemented using a Newton-cooling type algorithm that brings the plasma density, the internal energy and the momentum back towards their initial values on a time-scale that increases exponentially with distance from the top boundary. The damping time-scale is chosen such that a short wavelength amplitude pulse is fully damped on its way through this region (the characteristic Alfv{\'e}n crossing time). This approach was first used in \citet{2013ApJ...771...20M} where it proved to work excellent even in the situation where a high velocity jet impacts the top boundary for a long period of time. A side effect of the layer is that it acts as a sink/source for the plasma density and thermal energy when the region below the damping zone changes its characteristic parameters. 

\begin{table}
  \caption[]{Key parameters for the experiments presented in the
             paper. These represents the driving velocity, the scaling
             of the driving time and the hight of the domain relative the horizontal size. 
             A value of 0.1 for the driver velocity and 1.0 for the driving time imply realistic 
             values.} 
  \label{t1.tab}
  \begin{center}
  \begin{tabular}{c c c c}
      \hline
      \noalign{\smallskip}
      name & V amplitude & t drive & height \\
      \noalign{\smallskip}
      \hline
      \noalign{\smallskip}
      Exp-1\tablefootmark{a} & 0.1 & 1. & Half \\  % previous Exp-4, Correct combination of B and V
      Exp-2\tablefootmark{a} & 0.1 & 2. & Half \\
      Exp-3\tablefootmark{a} & 0.1 & 1. & Half \\
      Exp-4\tablefootmark{b} & 0.1 & 20.& Quarter \\
      Exp-5\tablefootmark{c} & 1.0 & 1. & Half \\  % previous Exp-1, wrong combination of B and V
      Exp-6\tablefootmark{c} & 0.1 & 1. & Half \\
      Exp-7\tablefootmark{c} & 0.1 & 10.& Half \\
      \noalign{\smallskip}
      \hline
   \end{tabular}
   \tablefoot{
      \tablefoottext{a}$-$ {potential model based on HMI data for the upflow event.}
      \tablefoottext{b}$-$ {as (a), but only using each 20th HMI image for determining the flow profile.}
      \tablefoottext{c}$-$ {potential model based on HMI data taken one day before the observed upflow.}
   }
\end{center}
\end{table}
%
%________________________________________________________________
\section{Results}
\label{res.sec}

This section discusses the general evolution of the experiments, using {\it Exp-1} as the reference case. Significant deviations found in the other experiments are discussed when appropriate. The time evolution of the general magnetic field topology, the identified velocity patterns, the typical Poynting flux, and the mass evolution of the active region are discussed below. 

% impact of the driving on the boundary motion
%________________________________________________________________
% Topology evolution
\subsection{Magnetic topology evolution}
\label{topology.sec}
To understand the dynamical evolution of the experiments an insight of the magnetic field topology is required. This provides information about the spacial locations where interaction between different magnetically connected domains may be able to drive systematic flows for longer periods of time. That is if the involved boundary regions are exposed to the correct systematic stress. Understanding the magnetic skeleton defining the topology of the magnetic field requires knowledge of the magnetic null point locations \citep{1996A&A...308..233B} and their local structure, spine axis and fan plane \citep{1996PhPl....3..759P}, the presence of separatrix lines connecting individual null points and the possible existence of QSLs \citep{1995JGR...10023443P, 2007ApJ...660..863T} that are independent of the null points. 
 
To investigate the skeleton, magnetic null points are identified using the method developed by \citet{2007PhPl...14h2107H}. This approach assumes the magnetic field to be well represented by a trilinear interpolation between the grid-points. For the initial magnetic field configuration, defined below, this assumption is only challenged close to the photospheric boundary where the fastest changes in the magnetic field exist. Null points found just a few grid points above this surface easily fulfils this requirement due to the exponential decay in complexity of the field with height.

\begin{figure}
   \centering
   \includegraphics[width=7cm]{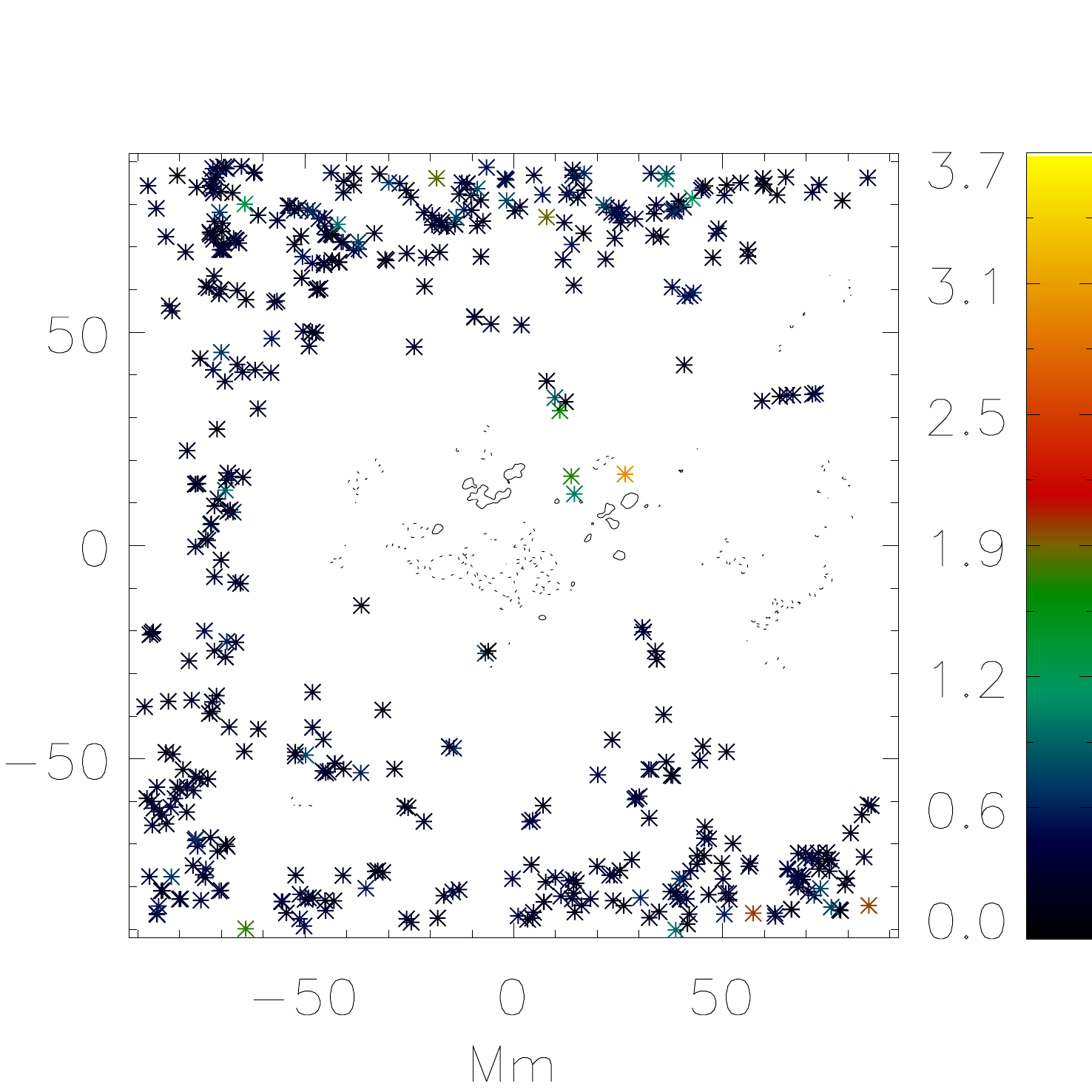}

   \includegraphics[width=7cm]{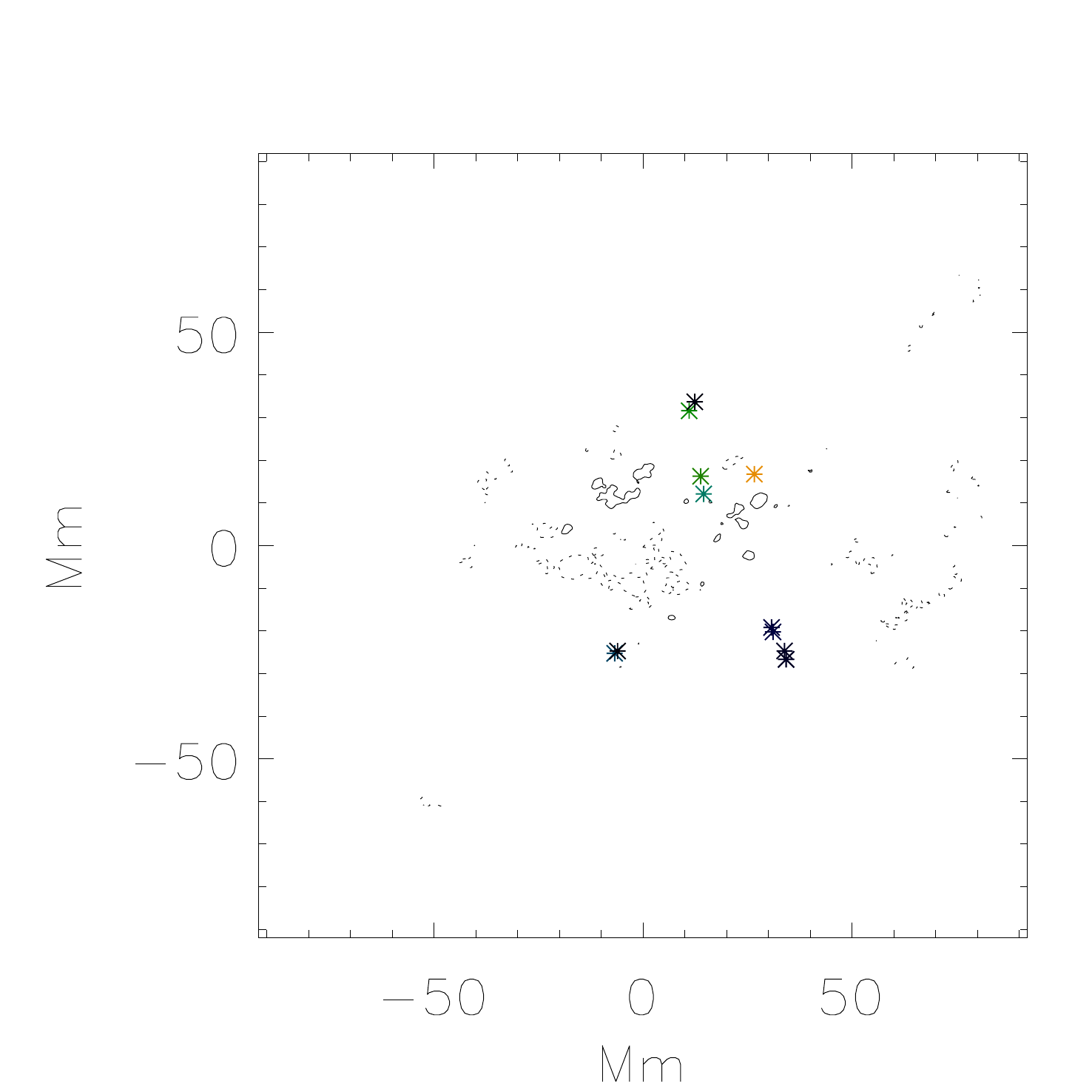} \,\,\,\,\,
   \caption{Horizontal positions of the null points in the initial
            potential magnetic field relative to the frame centre. 
            The colour coding of the points indicate the heights of the nulls points .
            The square-root of the height is shown to compress the scale. 
            The contour lines (full and dashed) show the locations of the strong
            magnetic flux concentrations ($\pm$500 G). 
            TOP: 74 null points above 0.54 Mm.
            BOTTOM: The 13 central null points listed in \Tab{t2.tab}. 
           }
   \label{nulls.fig}
\end{figure}

The initial potential magnetic field is extrapolated from the HMI data taken on November 13, at 13:02\, UT, as described above. The size of the domain is chosen such that the field line structure around the active region complex is not critically influenced by the imposed 2D periodic boundary conditions. In contrast to this the field lines connectivity close to the periodic boundaries does not represent the real solar magnetic field and a large fraction of these would most likely be closed over a larger distance. \Figu{nulls.fig} shows the horizontal positions of the null points in the domain that are located at least one grid point above the bottom boundary. Their colour indicates the height of the nulls. Two things are to be noticed. First, a large fraction of the nulls are scattered around the edges of the 2D horizontal domain, where they are introduced by the artificial damping of the normal component of the magnetic field, and also by the rapid changing of the weak field structure. These nulls are, therefore, not considered in the following investigation. Second, a number of nulls are located in the close vicinity of the active region. These will be discussed below, as they are important for understanding the topological structure of the magnetic field around the observed upflow region.

\begin{figure}
   \centering
   \includegraphics[width=8cm]{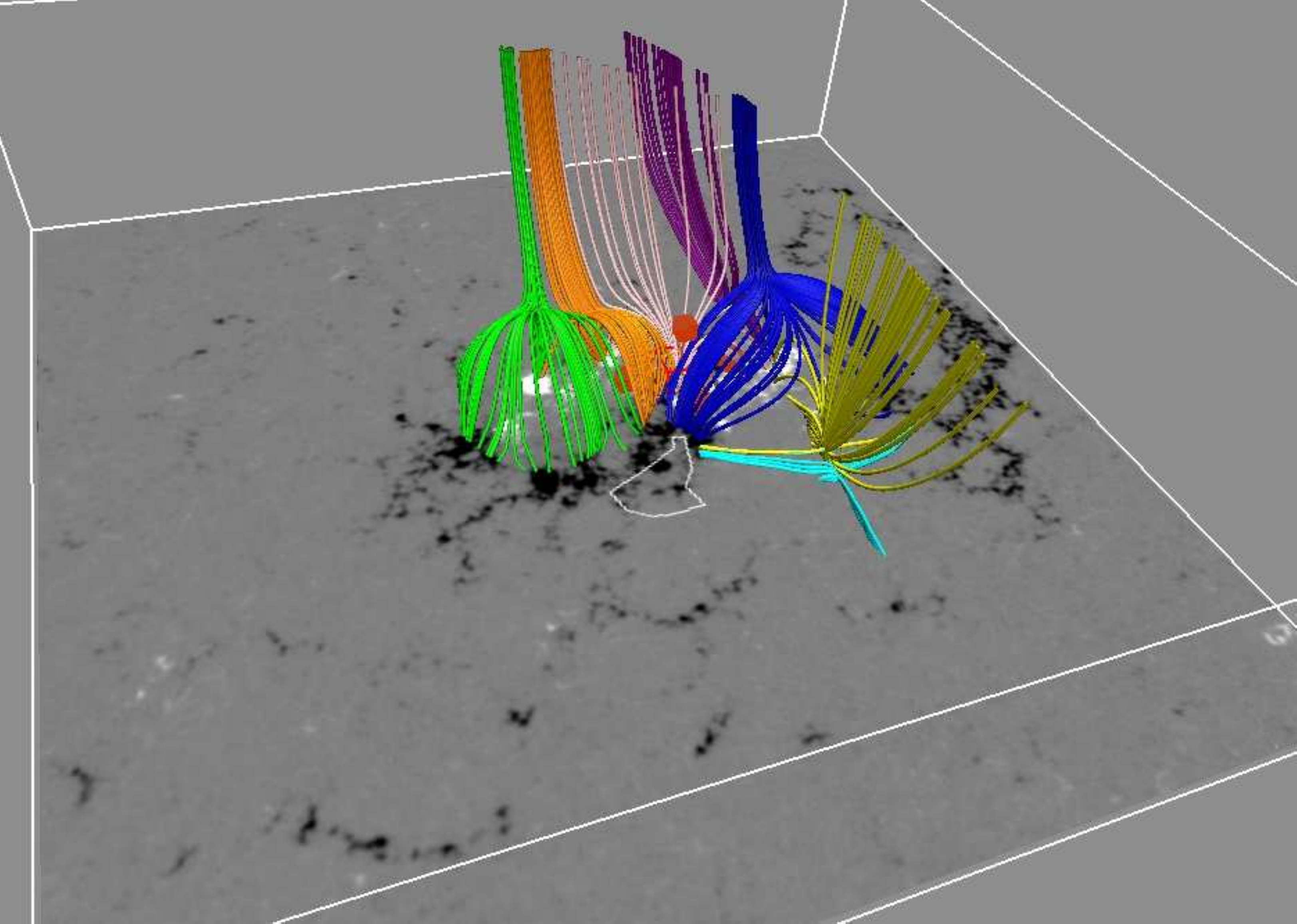}
   \caption{Representation of the initial magnetic field line topology. The different coloured 
            field lines are traced from the vicinity of different null points, above the active 
            region. These indicate the characteristic connectivity of the magnetic flux. The 
            spatial orientation and size are comparable to \Fig{fig1.fig}. The base is
            the real magnetogram scaled to $\pm$~500 G. The white contour line in front of the
            footpoints of the magnetic field lines represents the footprints of the outflow region 
            marked for a more detailed investigations in Fig.(6) in Paper~I.
           }
   \label{null_field.fig}
\end{figure}

\begin{figure}
   \centering 
   \includegraphics[width=8cm]{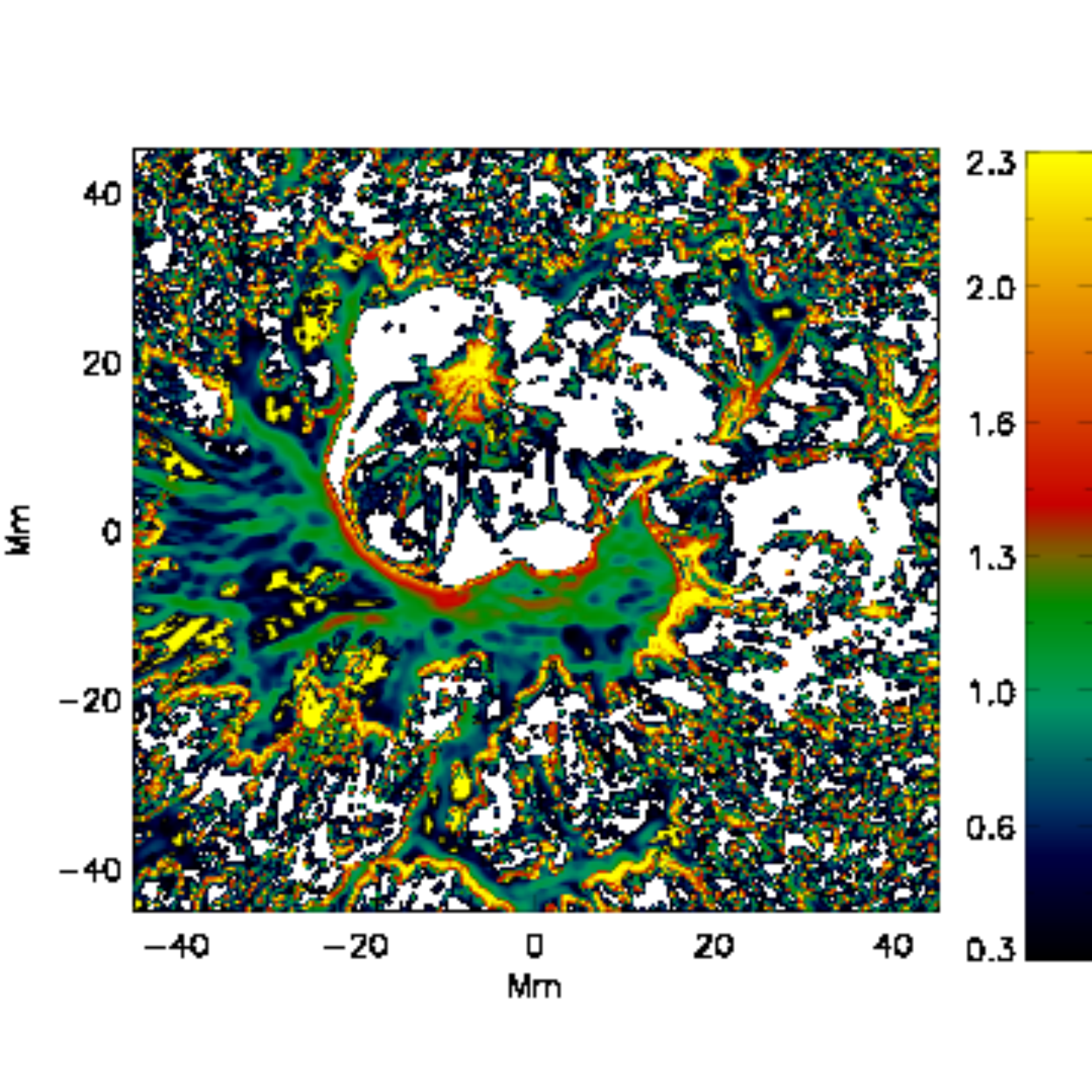}

   \includegraphics[width=8cm]{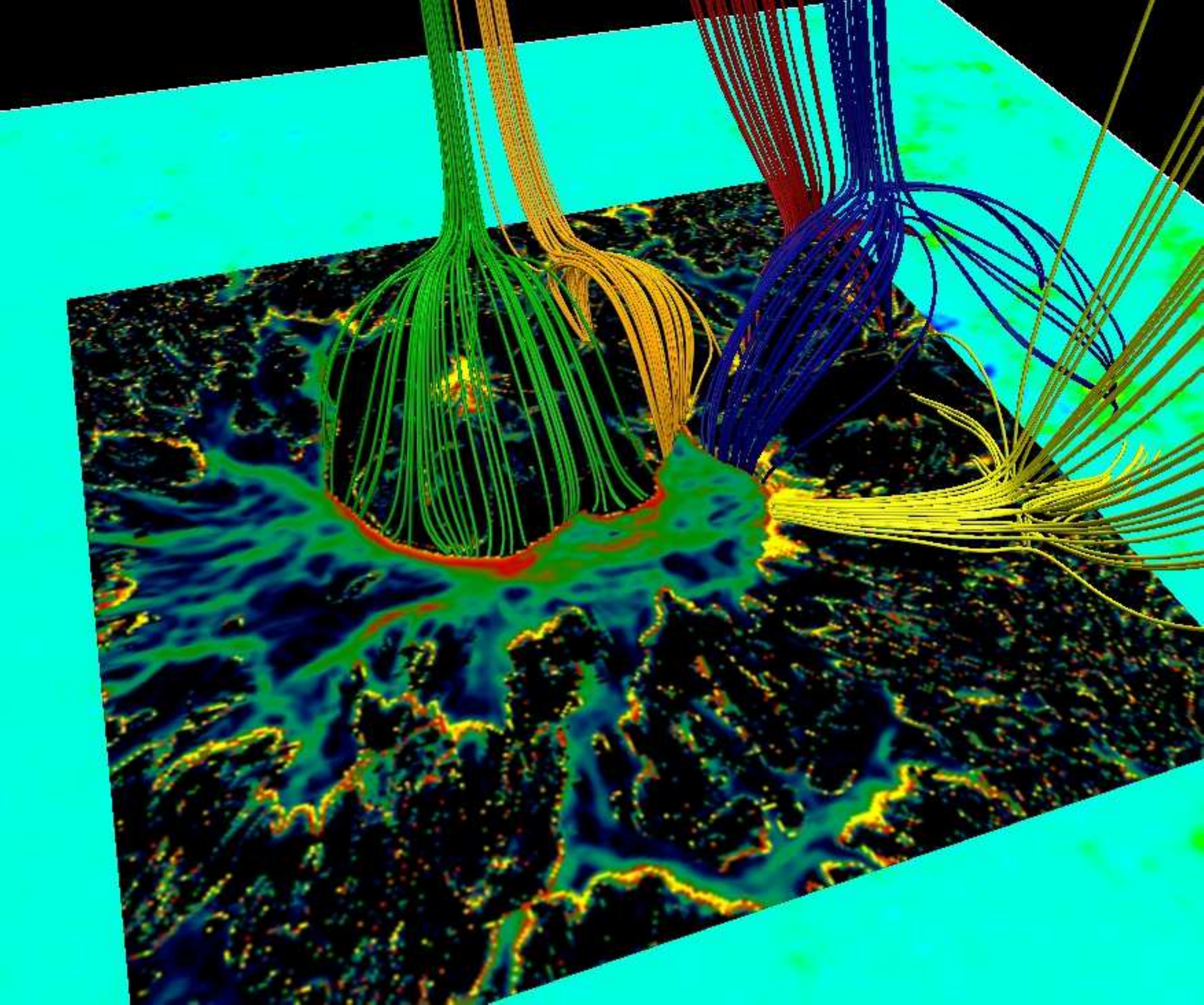}
   \caption{TOP: Quasi separator layer map indicating the locations of rapid changes in 
            the field line mapping across the core region of the domain using the invariant 
            measure defined by  \citet{2007ApJ...660..863T}. 
            The data represent $\pm 45 Mm$ in the $x$ and $y$ around the centre of the 
            domain. The data are logarithmically scaled over values from 2 to 1000. White areas 
            have low values.
            BOTTOM: QSL map and its relation to the null points close to the active region.
           }
   \label{QSL.fig}
\end{figure}

\Figu{null_field.fig} shows the skeleton of the magnetic field in the vicinity of the active region. The data slice in the bottom of the frame represents the magnetogram on to which the region of the footpoints of the upflow defined in Paper~I is marked with a white line.  
The different coloured field lines represent the structure of the magnetic field close to the most important null points, clearly showing the spine axis and fan planes that separate the magnetic field line connectivity into independent flux regions. 
The nulls represented by blue and green field lines both have their fan planes extending down towards the photosphere, rooted in the negative (red) polarity flux. The spine axis of these nulls are, therefore, connecting either to the  positive (purple) flux in the photosphere or to the top boundary of the domain. The three other coloured field line regions are more complicated and can, from \Tab{t2.tab}, be seen to contain more than one null point. The spine axis of the dominating null points (yellow and red) are typically in the horizontal direction, which is clearly seen for the yellow null region. The fan planes of these nulls, on one side, reach towards the spine axis of the blue and green nulls, and therefore intersect with their fan plane. This implies that several separator lines connecting between the different nulls, providing the division of the magnetic field line connectivity into several independent flux domains. 
Depending on where the field lines are traced from the photosphere they either belong to a closed or an open loop system with regard to the extrapolation of the limited field region. Especially, it is noticed that all field lines starting from the negative flux in the active region are closed field lines connecting to the positive flux regions due to the general excess of positive flux in the active region. The negative flux regions therefore contain both closed and open field lines, where the open field lines connect mainly to the top boundary of the domain. This is seen in the bottom frame of \Fig{fig1.fig}, where the red field lines outside the null dome regions connect from the photosphere to the top boundary. This open field line structure is an artefact of the limited region used for the field extrapolation. The position of the thirteen null points are given in \Tab{t2.tab} for the initial condition in {\it Exp-1}. 
Investigations of magnetic field extrapolations have often used the concept of QSL 
%\citep{1995JGR...10023443P, 2007ApJ...660..863T} 
to identity locations in space where the field line mapping changes particularly fast and reconnection therefore may be initiated. A QSL does not need to contain a mathematical discontinuity in the field line mapping, as implied by the presence of 3D null points, while a null point can be seen to be embedded in a QSL region \citep{2013ApJ...774..154P}. Quasi separator layer regions, like null points, fan planes and separator lines, are found to be locations in space where current easily accumulates when the magnetic field configuration is appropriately stressed. 
\Figu{QSL.fig} shows a QSL map for the central region of the domain. It is seen that the main QSL regions are located around the locations where the fan planes of the overlying null points intersect the photosphere. Other QSL regions are also seen further away from these null defined regions. These may be associated with the many null points located closer to the domain boundary (see top frame of \Fig{nulls.fig}). This indicates that the magnetic field contain lots of small-scale variations due to the actual mapping outside the active region. It is clear from the analysis of the magnetic skeleton that the underlying structure is very complex, which allows many different reactions to the imposed boundary driving.

\begin{table}
  \caption[]{Positions of the 13 null points embedded close to the active region 
             (in Mm measured from the centre of the domain ($x,y$) and hight above the
             photosphere ($z$)).}
  \label{t2.tab}
  \begin{center}
  \begin{tabular}{r r r r}
      \hline
      \noalign{\smallskip}
      \# & x & y & z \\ 
      \noalign{\smallskip}
      \hline
      \noalign{\smallskip} 
%       1 &      35.4265 &     -29.2866 &    0.0311783 \\
%       2 &      36.0895 &     -22.4961 &     0.642942 \\
%       3 &      32.6493 &     -20.6628 &    0.0124169 \\
%       4 &      31.8198 &     -19.1594 &    0.0650185 \\
%       5 &      36.0465 &     -17.0905 &    0.0696568 \\
%       6 &      36.1284 &     -17.1809 &      2.70726 \\
%       7 &      18.8320 &      10.4018 &     0.755203 \\
%       8 &      37.8525 &      18.7167 &      14.0516 \\
%       9 &      41.1355 &      25.2161 &    0.0446176 \\
%      10 &     -3.01579 &      28.4939 &      6.49838 \\
%      11 &      37.5709 &      29.1313 &      1.42699 \\
%      12 &      36.7692 &      31.0756 &     0.335421 \\
%      13 &     -5.35785 &      32.1051 &     0.109239 \\
       1 &      35.4 &     -29.3 &      0.03 \\
       2 &      36.1 &     -22.5 &      0.6 \\
       3 &      32.6 &     -20.7 &      0.01 \\
       4 &      31.8 &     -19.2 &      0.07 \\
       5 &      36.0 &     -17.1 &      0.07 \\
       6 &      36.1 &     -17.1 &      2.7 \\
       7 &      18.8 &      10.4 &      0.8 \\
       8 &      37.9 &      18.7 &     14.1 \\
       9 &      41.1 &      25.2 &      0.04 \\
      10 &      -3.0 &      28.5 &      6.5 \\
      11 &      37.6 &      29.1 &      1.4 \\
      12 &      36.8 &      31.1 &      0.31 \\
      13 &      -5.3 &      32.1 &      0.1 \\
      \noalign{\smallskip}
      \hline
   \end{tabular}
\end{center}
\end{table}       


\begin{figure}
   \centering 
   \includegraphics[width=7cm]{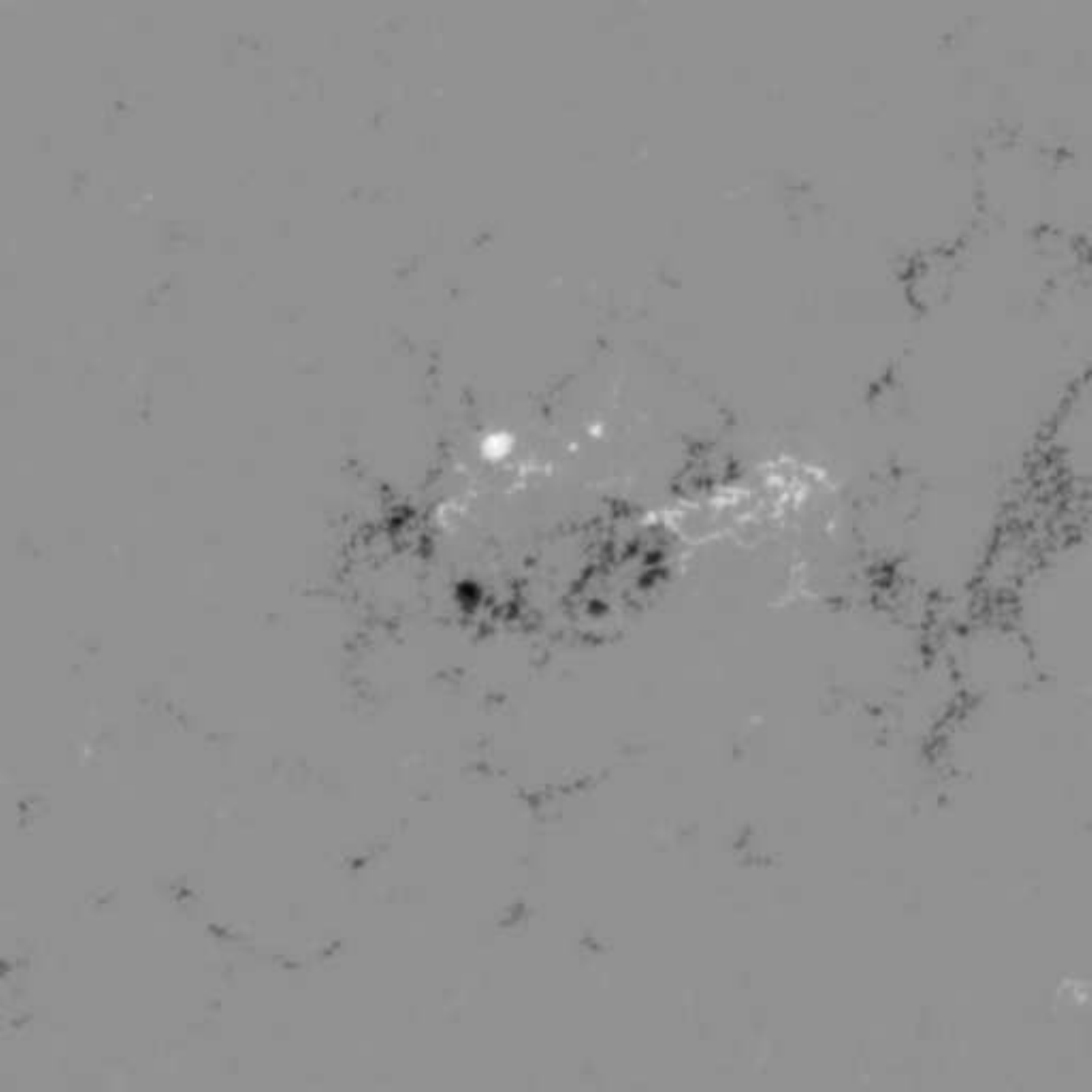} 

   \includegraphics[width=7cm]{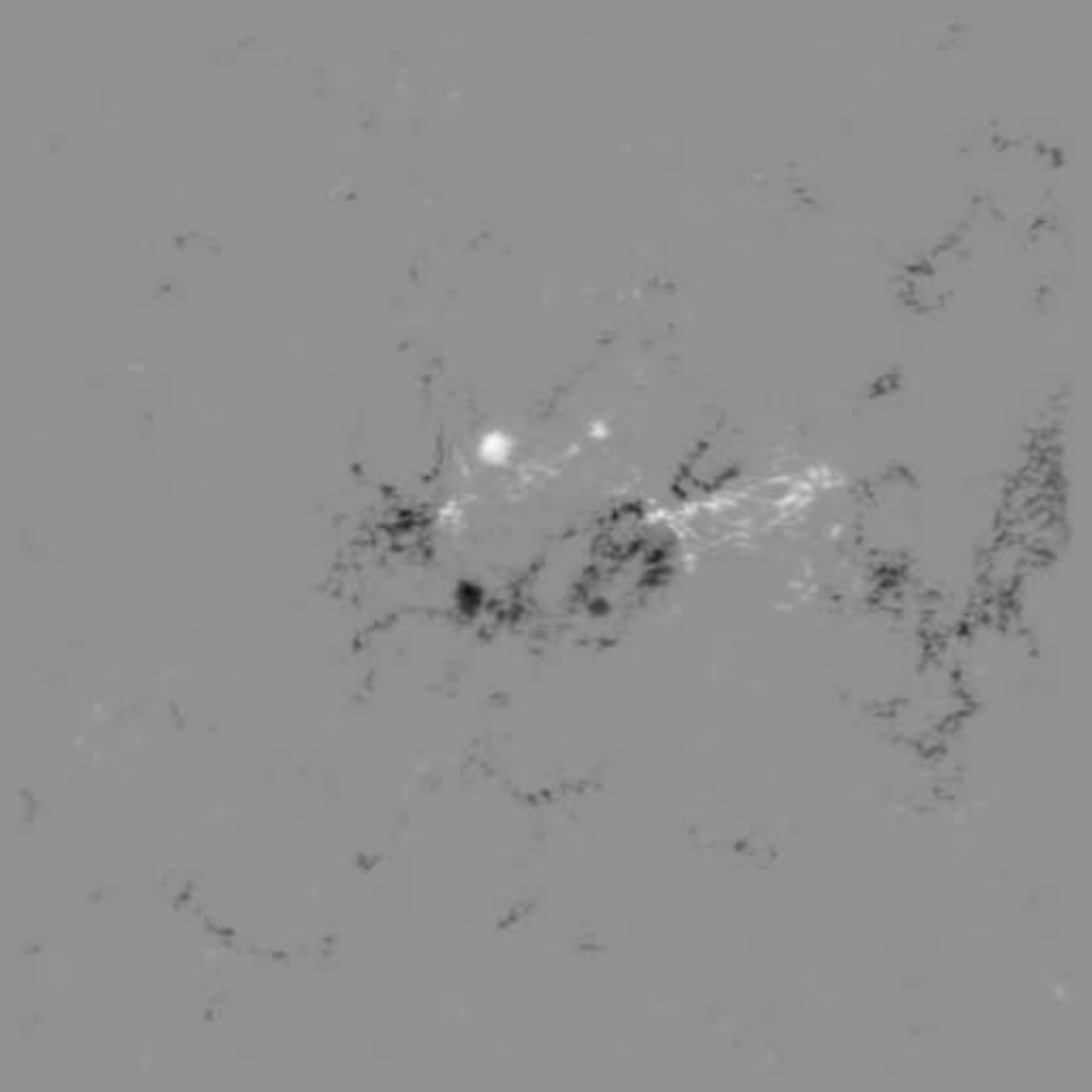}
   \caption{Comparison of the magnetograms from the HMI observations and the numerical experiment. 
            The top frame represents HMI data 193.5 min after the initial state seen in the top
            frame of \Fig{fig1.fig}. The bottom panel represents the flux distribution 
            after the same time derived by the MHD code. The dynamical scale in the frames 
            represent $\pm1500$ G.
           }
   \label{driver_compare.fig}
\end{figure}

The imposed driving does not introduce large changes to the flux distribution on the model photosphere over the duration of the experiment and it does not allow for flux to emerge or submerge. This is both an advantage and a limitation with the imposed flow pattern and the general handling of the boundary conditions. Comparisons between the HMI observations and the evolved model show clear differences and that these increase with time. An example is shown in \Fig{driver_compare.fig}, that represents one time in the evolution. A more detailed comparison shows that most of the differences are due to small scale dynamical evolution in the observed magnetograms, that is not represented by the driver obtained from the LCT. 

%________________________________________________________________
%   velocity evolution 
\subsection{Velocity evolution}
\label{velocity.sec}
%Ideally we are looking for an explanation for the consistent outflow found in the observations of a number of active regions \citep{Baker2009, Harra2008, 2008ApJ...685.1262M, vanDriel-Gesztelyi2012, 2010SoPh..261..253M, ...}. 
To identify the mechanism behind the upflows, regions with systematic positive vertical flow velocities need to be located in the experiment. For this, three different visualisations of the data are used. 

\begin{figure}
   \centering
   \includegraphics[width=7cm]{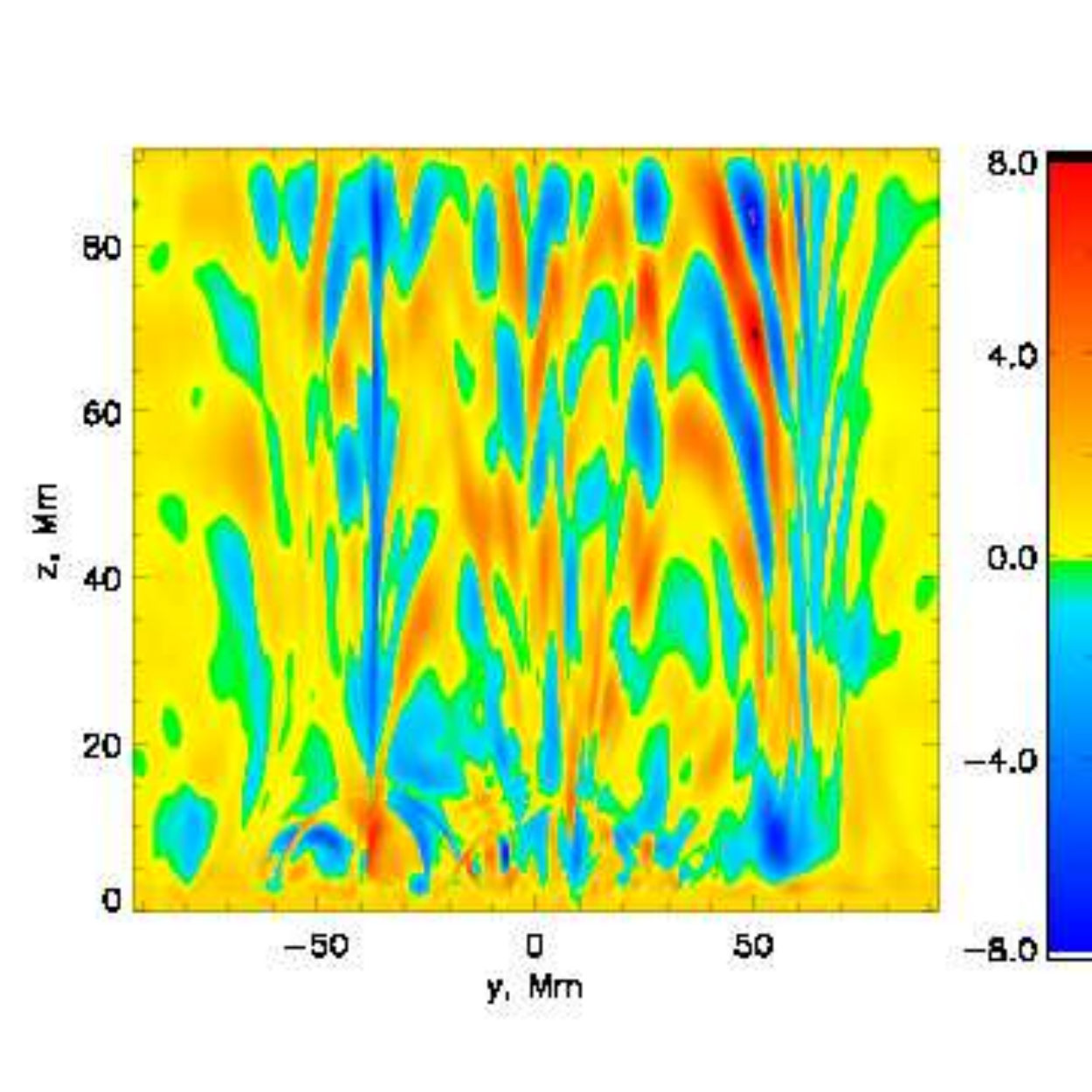}

\vspace{-1cm}

   \includegraphics[width=7cm]{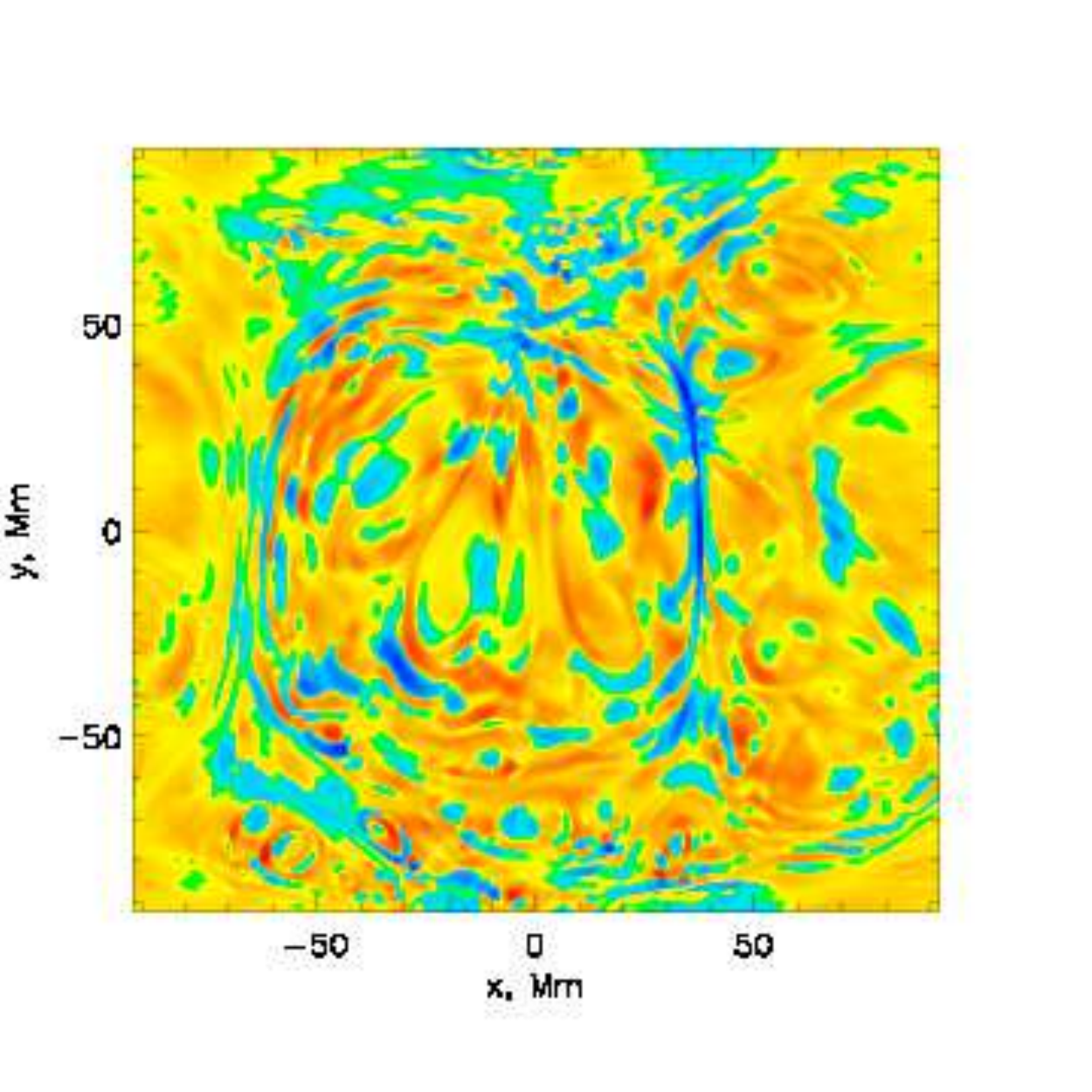}  

\vspace{-1cm}

   \includegraphics[width=7cm]{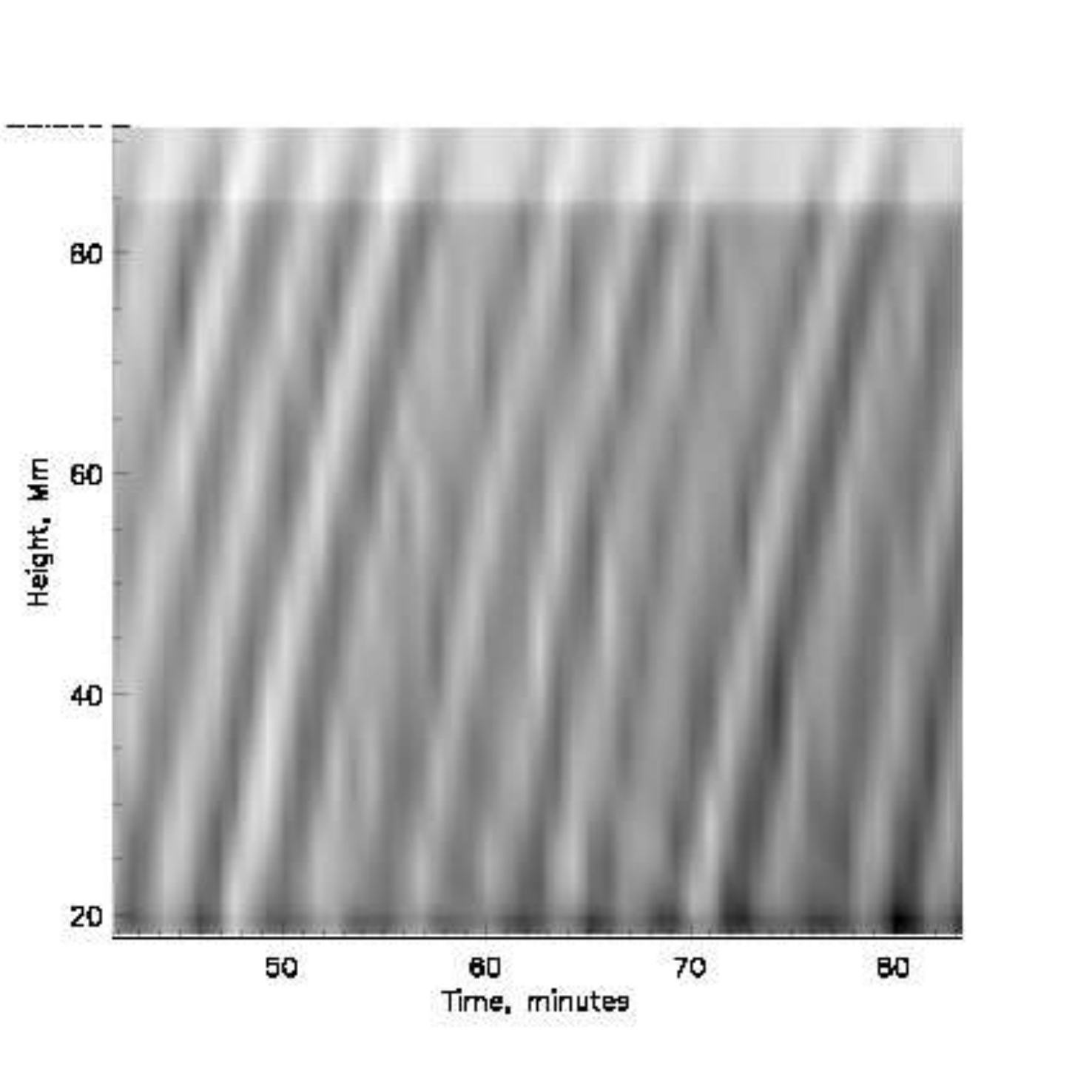}
   \caption{ TOP: Representation of a vertical slice (y,z-plane) in 
             the observed upflow region of the domain, 
             showing the vertical flow speeds of $\pm$8~\kms. % 0.8 code unit.
             MIDDLE: The same but for a horizontal plane (x,y plane inside the corona). 
             BOTTOM: Time slice of the density variation in height, showing 
             how structures are moving upward in the atmosphere with a speed 
             of roughly 150~\kms.  %15 code units.
             The plot also show the period of the oscillations is on the order of
             3.7 minutes. 
             The snapshots used for the top and middle frames represent t=132.5 min. %t=79.5 
            % in code units. 
            }
   \label{flow_speed.fig}
\end{figure}

First, we investigate the vertical velocity in a 2D vertical plane that in its horizontal direction passes through the upflow region identified by the observations (north-south orientation). This shows complicated flow patterns covering the whole plane, with an alternating velocity structure that is advected upwards in time. An example of this is shown in the top panel of \Fig{flow_speed.fig}. This flow pattern represents velocities with a peak of the order 10~\kms. The alternating flow patterns are wave motions that represent internal compression/decompression speeds in the vertical direction, while the wave patterns move upward along the field lines with a much higher speed. 

Second, by only investigating a single vertical plane one can easily miss different flow behaviours in other parts of the domain. This can be helped by investigating the vertical flow speed in a horizontal plane, at the base of the coronal region, where one can easily look for systematic flow patterns. An example is shown in the middle panel of \Fig{flow_speed.fig}. This shows the whole plane is covered with alternating velocity patterns. Following these structures in time, the patterns are found to be constantly changing without showing any localised regions containing a systematic flow pattern.

Third, to get an estimate of the upward motion of the flow patterns, one can plot the relative changes in the plasma density from a single column in height as a function of time. This shows the upward propagation of density structures. The inclination of the patterns represents the propagation speed, bottom panel of \Fig{flow_speed.fig}. Using this to estimate the propagation velocity in the coronal part of the domain, between 20 and 80 Mm in height, one gets a velocity of the order 150~\kms. This simple visualisation does not take into account the magnetic field line structure, but at this hight in the atmosphere its direction is dominantly in the vertical direction.

\begin{figure}
   \centering
   \includegraphics[width=8cm]{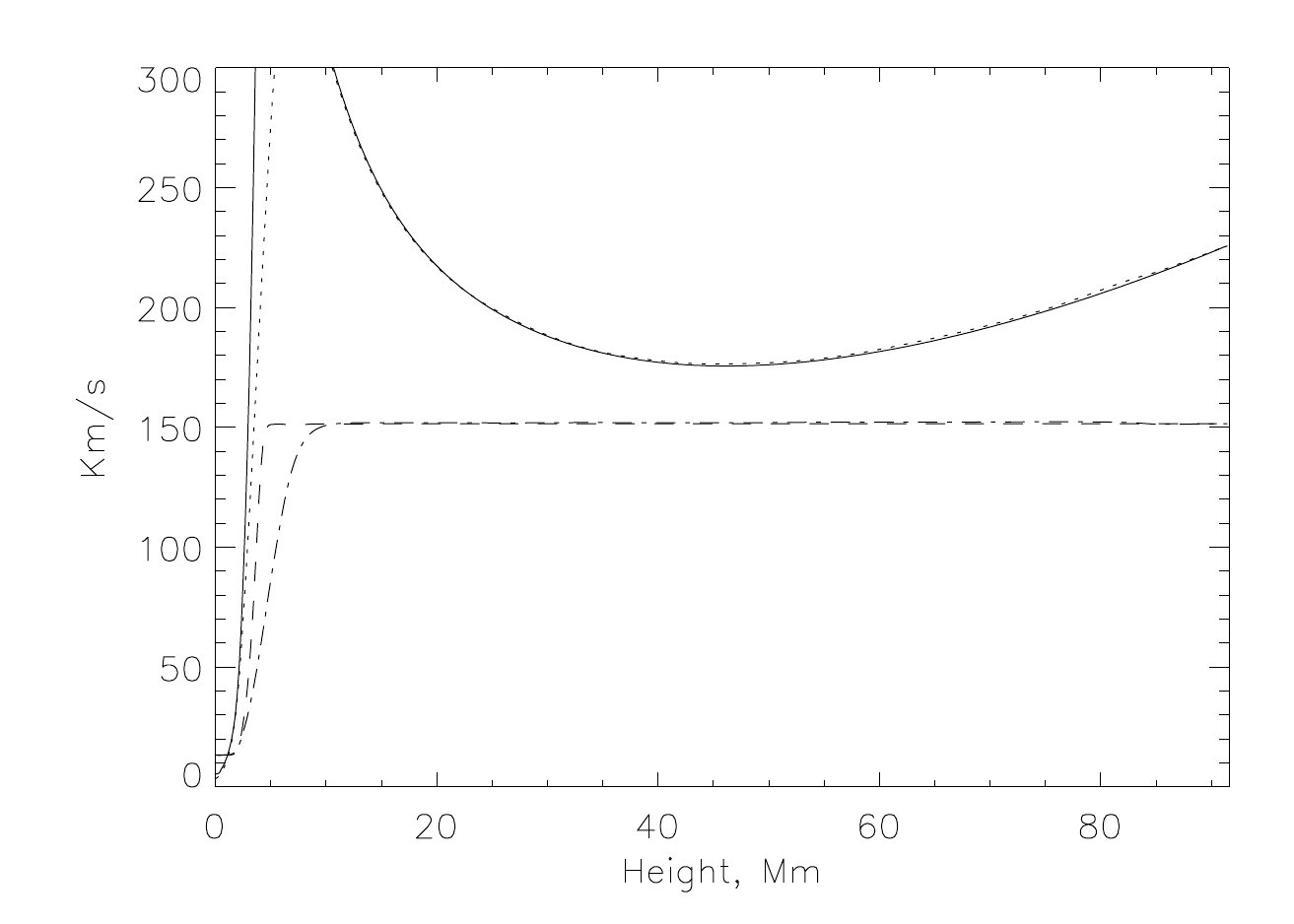}
   \caption{Characteristic speeds of the Alfv{\'e}n velocity and the sound speed. 
            The full line shows the Alfv{\'e}n velocity for the initial
            setup. The dotted line shows the comparable profile at t=132.5 minute. %$t=79.5$.
            The same is shown for the sound speed with the dashed line showing
            the initial profile while the dot-dashed represents the later time.
           }
   \label{flow_speeds_model.fig}
\end{figure}

To check the speed of the upward propagating oscillations with the characteristics flow speeds in the atmosphere, \Fig{flow_speeds_model.fig} shows the Alfv{\'e}n and sound speeds for t=0 and t=132.5 min. % t=79.5. 
The overall speed scales between 0 and 485~\kms. The typical sound speed is 150--170~\kms\ in the coronal region, while the Alfv{\'e}n velocity is higher. 
This shows that a propagating wave in this coronal model will obtain rather different speeds depending on where in the domain it is located and which wave type it represents. The typical wave speed observed in the density perturbation shown in \Fig{flow_speed.fig} indicates a flow speed of around 150~\kms, which strongly indicates the waves to be a slow magnetosonic wave with a speed close to the sound speed. 

\begin{figure}
   \centering
   \includegraphics[width=8cm]{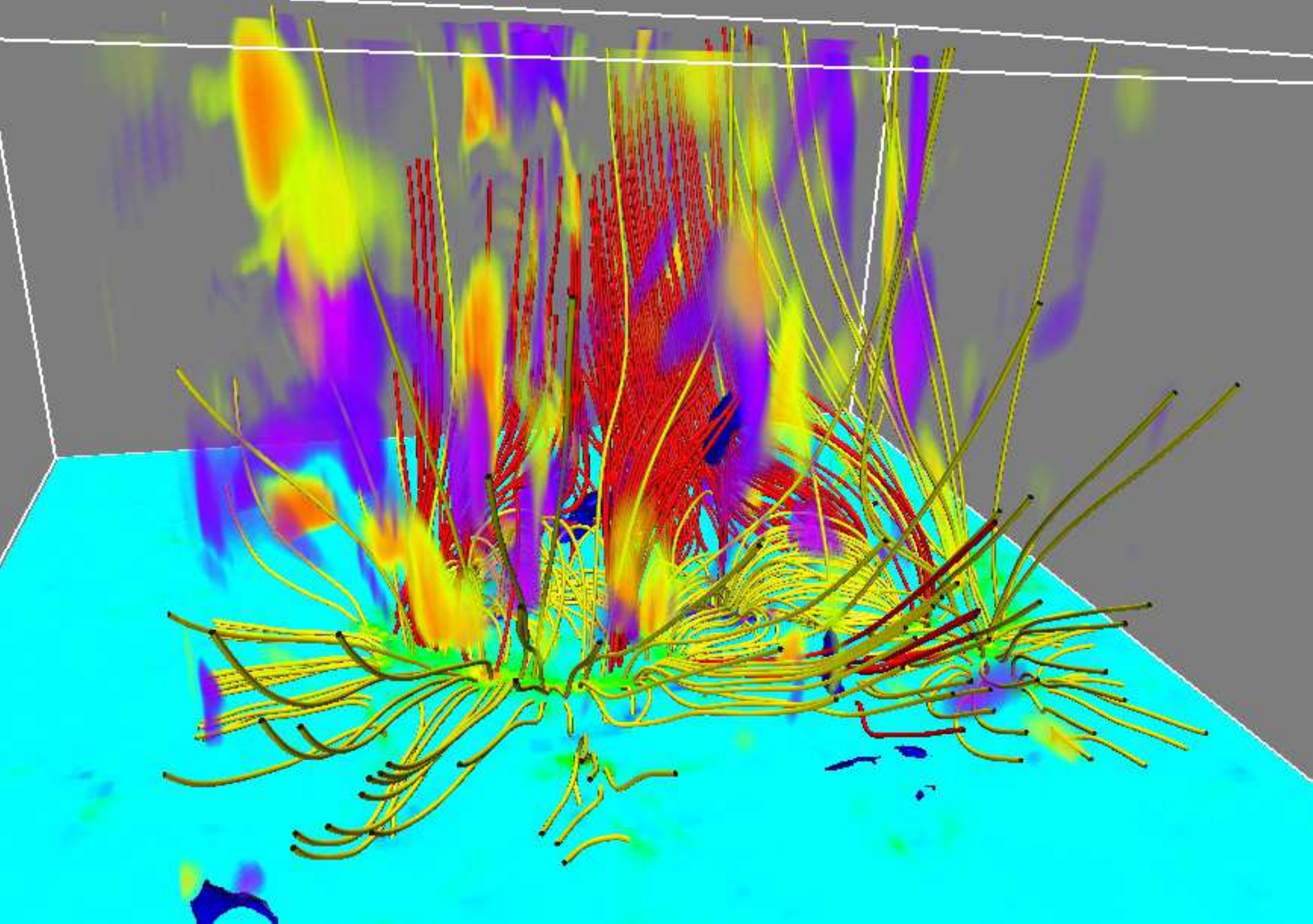}
   \caption{The magnetic field line structure is shown in the same way as in
            \Fig{null_field.fig}. Additionally, the red/purple blobs represent
            the vertical down/up flow velocities indicating the compression/decompression 
            patterns of the propagating waves.
            (See the attached movie for a time-dependent evolution of this visualisation, 
             it represents the time period from 83.3 - 325.0 min.)
           }
   \label{flow_speed_vapor.fig}
\end{figure}

\Figu{flow_speed_vapor.fig} shows a 3D representation of the field line structure combined with blobs of the vertical positive/negative flow velocity. Inspecting such visualisations of the data shows the oscillating flow patterns to be space filling and upwards propagating in time.

Repeating the experiment with different driving parameters (\Tab{t1.tab}), one finds the same characteristic behaviour, showing that the typical dynamical evolution is independent of the imposed driving velocity and characteristic timescale of the driver, and even of the driving pattern. The change in the driving time influences the characteristic timescale of the oscillations and wavelength of the travelling perturbations. 

% Average downflow
%  Exp-1      Exp-2
% -0.0185    -0.0305
% -0.0206    -0.0337
% -0.0232    -0.0384

%________________________________________________________________
% Poynting flux evolution
\subsection{Poynting flux evolution}
\label{poynt.sec}
\begin{figure}
   \centering
   \includegraphics[width=7cm]{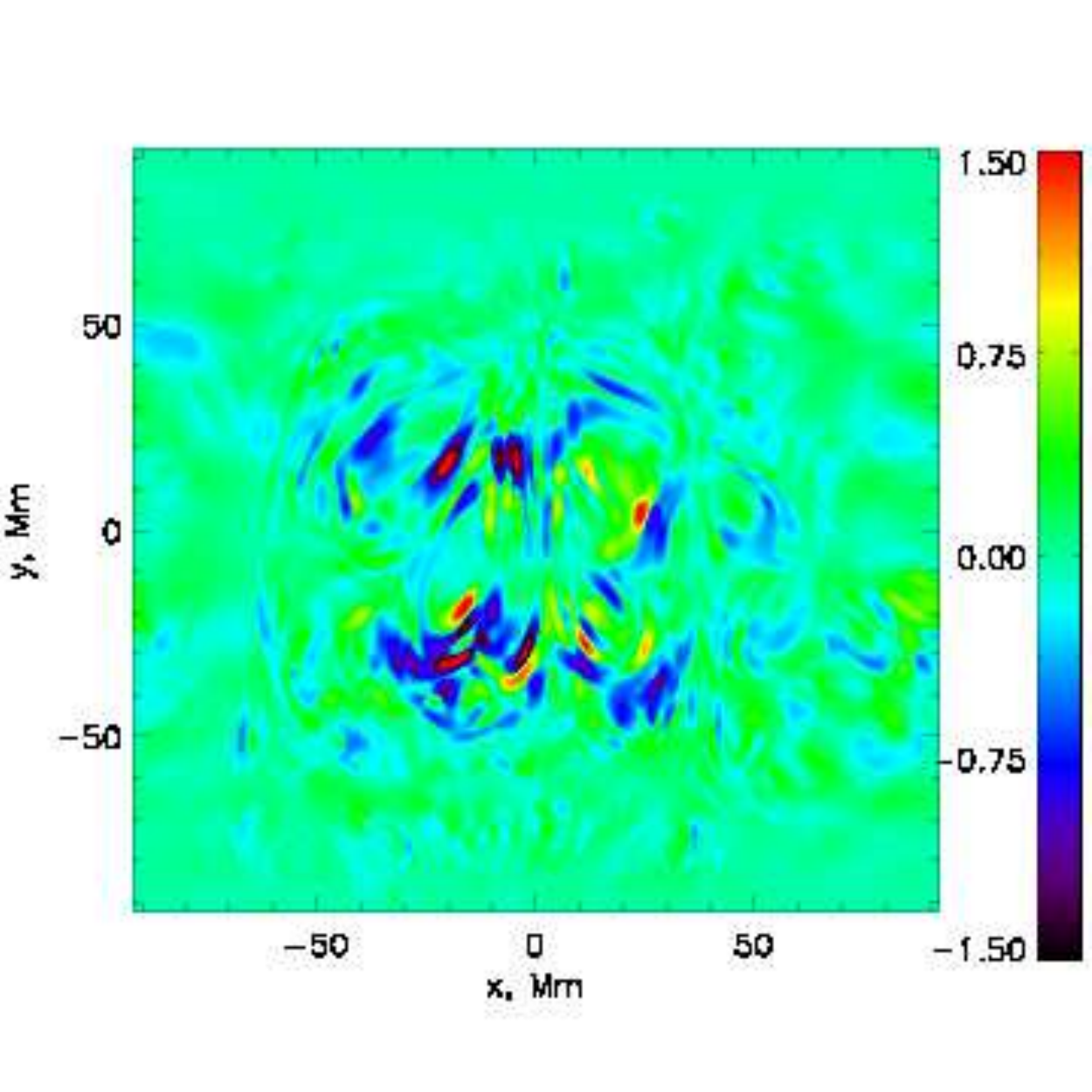} 

   \includegraphics[width=7cm]{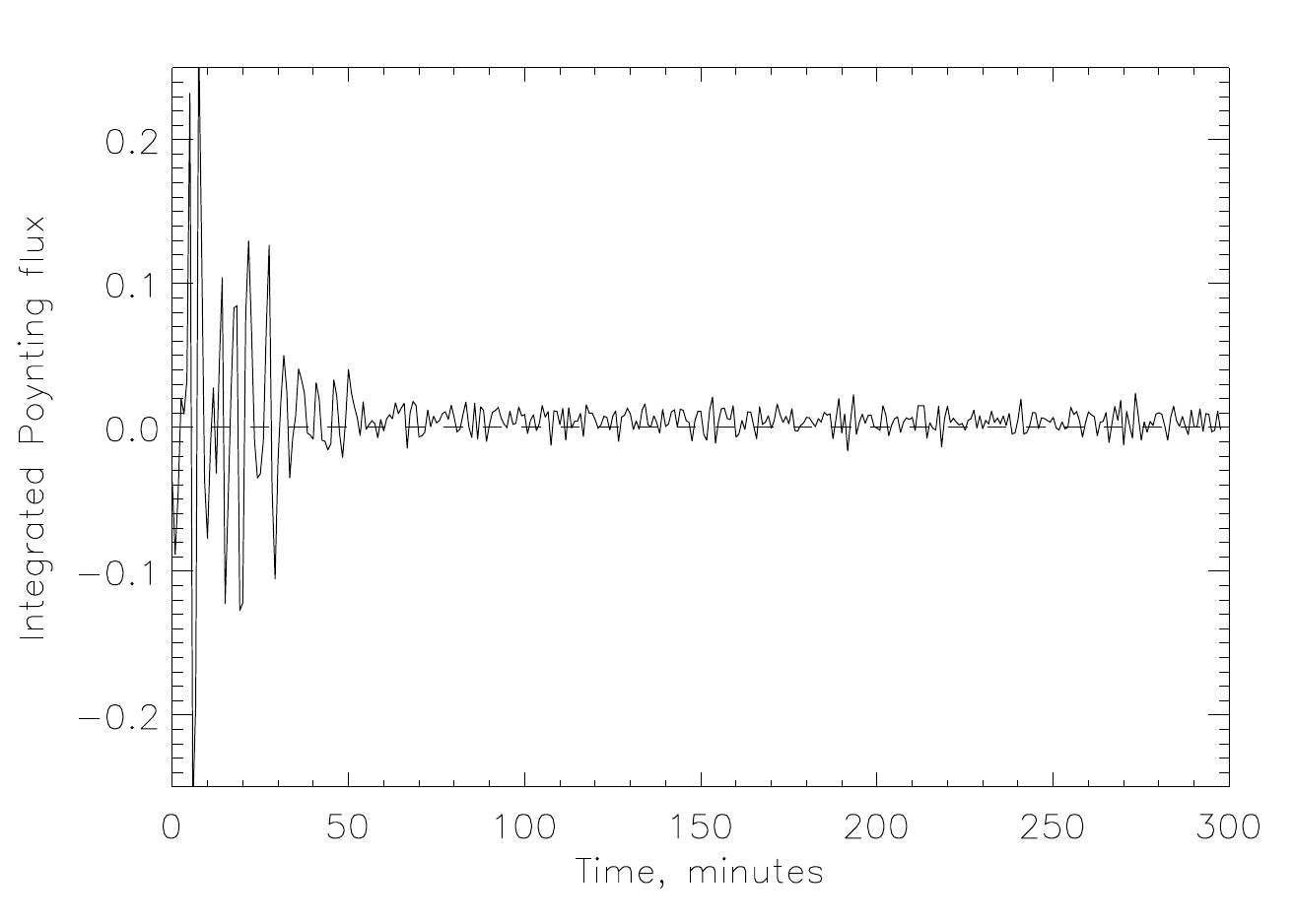} 
   \caption{TOP: Snapshots of the Poynting flux at a constant height in 
            the corona. The plane is from t=107.5 minute at a height of 35.9 Mm.
            BOTTOM: Time variation of the surface integrated Poynting flux
            as a function of time for the same height in the model.
           }
   \label{poynt_flux.fig}
\end{figure}

To measure the magnetic energy flux through different heights of the corona the Poynting flux is calculated:
\EQ
Pf(z_{const}) = (-\VV_z(\BB_x^2 + \BB_y^2) + \BB_z(\VV_x \BB_x + \VV_y \BB_y))/\mu_0.
\EN
The first contribution represents the advection of magnetic energy through the surface, while the second term is the work done by advecting the field lines in the horizontal direction in the plane. Time series of $Pf$ for different heights in the corona show that the time resolution of the saved data from the experiments is too low to directly follow the evolution of the Poynting flux patterns in the surface, this was also seen for the velocity components, and especially the horizontal components. It is clear that the patterns fluctuate significantly in both space and time. Initially this fluctuation is mainly concentrated above the active region, while the influence spreads to cover the whole plane as time progresses. 
To smooth out the influence of the low time resolution in the time series (50 seconds between data sets), one can instead look at the surface integrated Poynting flux as a function of time. This quantity indicates whether magnetic flux, on average, is transported through the plane as a function of time. The result is shown in \Fig{poynt_flux.fig}, where the top frame shows a snapshot of the Poynting flux through a vertical plane and the bottom frame shows the integrated Poynting flux through the plane as a function of time. This shows a small positive integrated Poynting flux through this plane that is about 2 orders of magnitude less than the peak values, indicating that magnetic energy, on average, is advected up through the plane to the volume above. Compared to the existing magnetic energy above this hight, the magnetic energy input is totally insignificant for the evolution of the magnetic energy over the time period covered by the experiment.
%The values in the left frame of \Fig{poynt_flux.fig} contain values within the range of $\pm$
%5e-5$ code units!!.
%0.063  ??
%The average of the plane is two order of magnitude less than the peak values, and has a value that is systematically below zero, indicating a very slow migration of magnetic energy downwards.  indicating that there is a good balance of the magnetic flux over both space and in time. 
Driving with a slower speed shows comparable spatial and time structures of the Poynting flux, with the space and time variations being even closer to zero.

%______________________________driver_compare.fig__________________________________
% Mass evolution
\subsection{Mass evolution}
\label{mass.sec}

\begin{figure}
   \centering
   \includegraphics[width=7cm]{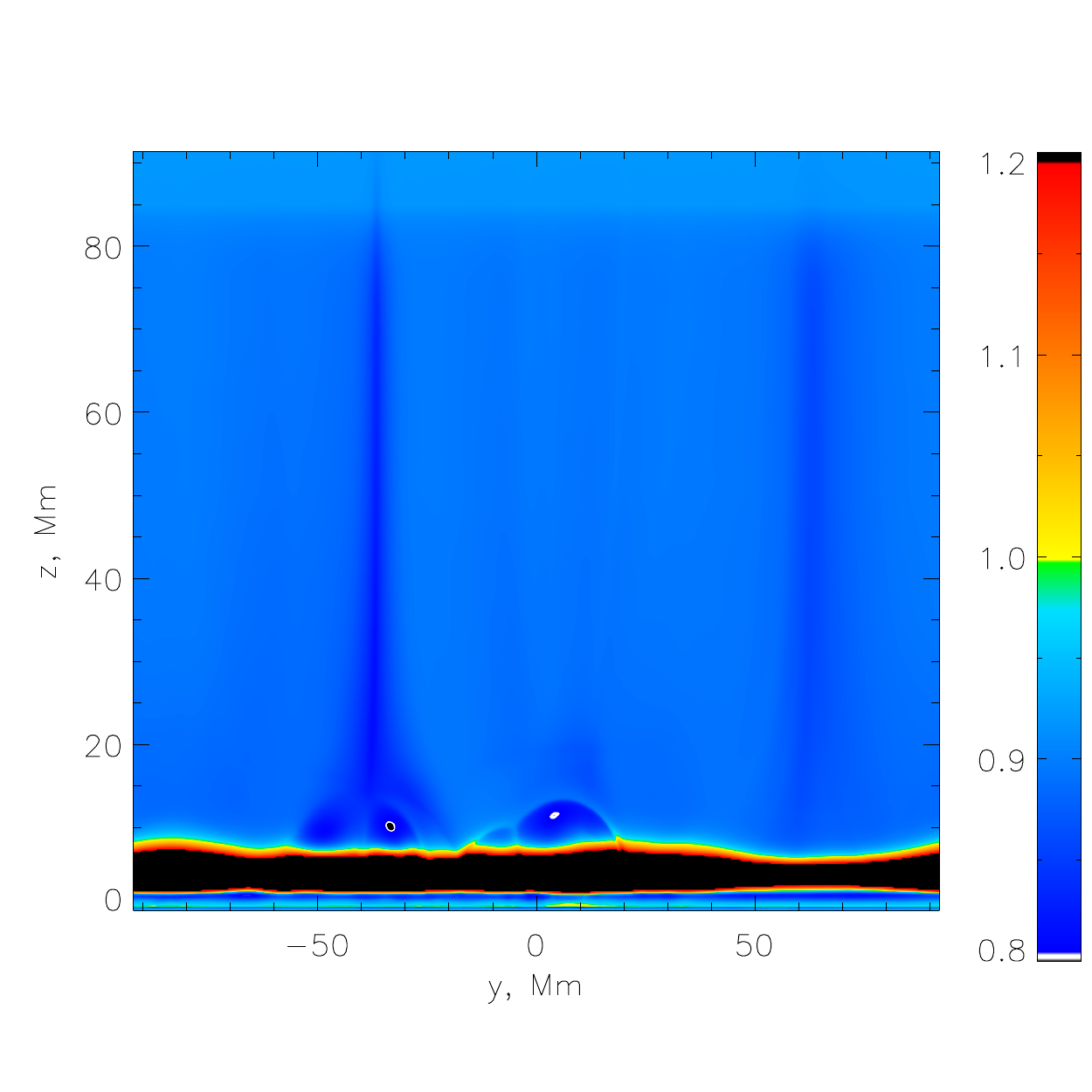}
   \caption{The percentage change of the density profile in a vertical plane between 
            80\% and 120 \% of the initial value.
            The plane is the same as shown in \Fig{flow_speed.fig}.
           }
   \label{density.fig}
\end{figure}

\begin{figure}
   \centering
   \includegraphics[width=8cm]{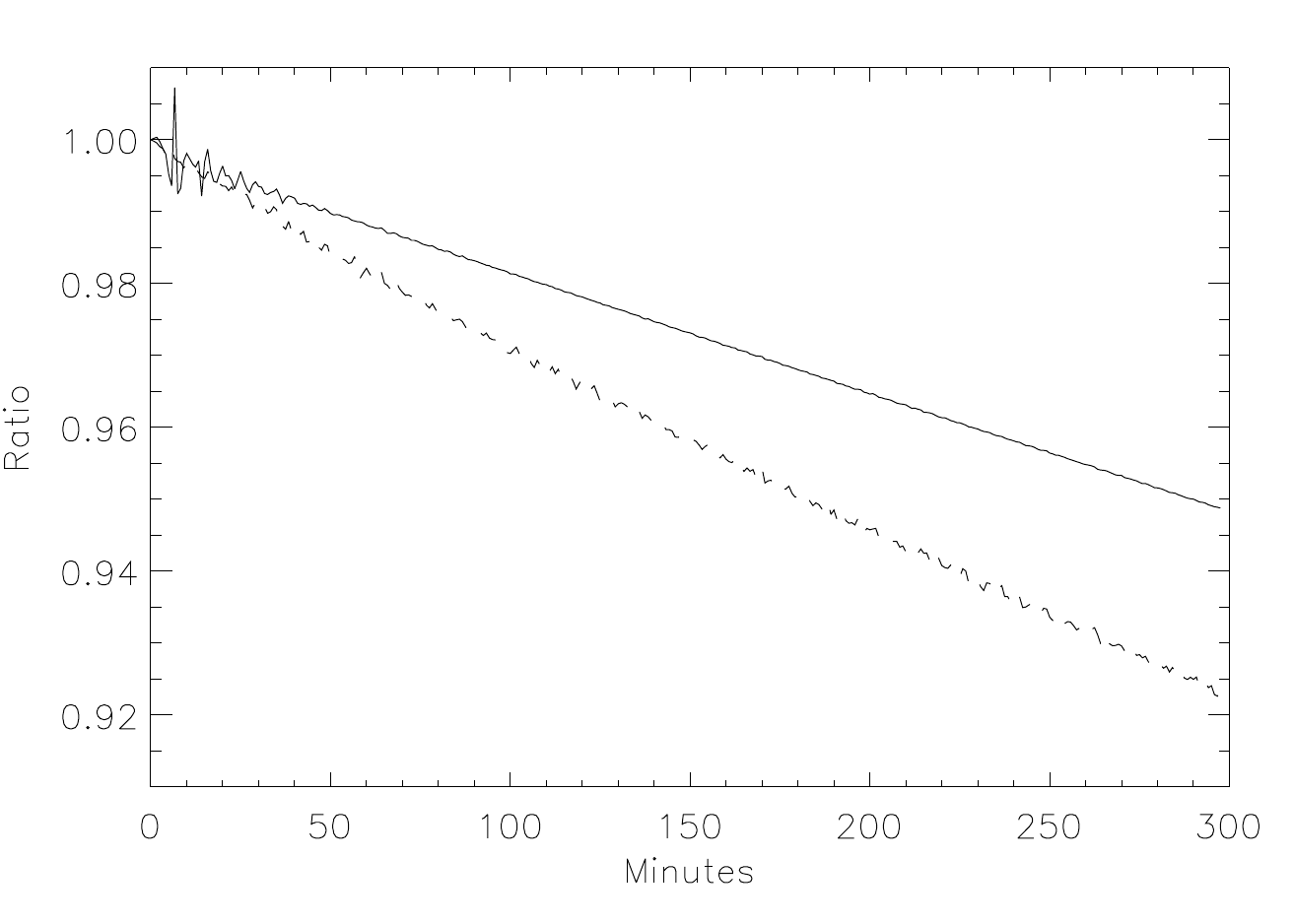}
   \caption{The time variation of the density of a plane at a constant height
            in the corona -- data from different heights in the corona show 
            the same time evolution. The full line represents experiment {\it Exp-1}, 
            and the dashed line {\it Exp-2}.
           }
   \label{density_time.fig}
\end{figure}

The initial density profile is shown as part of \Fig{fig2.fig}. Due to the large variations with hight, the simplest way to quantify density changes is to measure it relative to the initial profile. Doing this, two characteristic changes are seen in the data. First, it is found that the density of the medium transition region (height $3-6$ Mm) is nearly doubled as a time average of the full experiment. This significant increase in mass arises from two regions, one is mainly from an expansion of the layer below 3 Mm. This is caused by the imposed boundary stress which locally increases the magnetic and gas pressure of this region, making it expand in height. The second is a minor contribution from a general decrease in the density from the region above 6 Mm. The coronal mass is slowly being depleted. Large local variations from this behaviour is seen, showing fluctuation in both space and time. These variations depend on both the magnetic field line structure and the imposed flow profile. An example is shown in \Fig{density.fig} where the percentage change of the density for a vertical plane is shown. In \Fig{density_time.fig} the average percentage change for a horizontal plane is presented for both {\it Exp-1} and {\it Exp-2}. This shows a constant decrease of the mass in the coronal region, with the actual rate being dependent on the speed of the imposed driving velocity. 


\begin{figure}
   \centering
   \includegraphics[width=7cm]{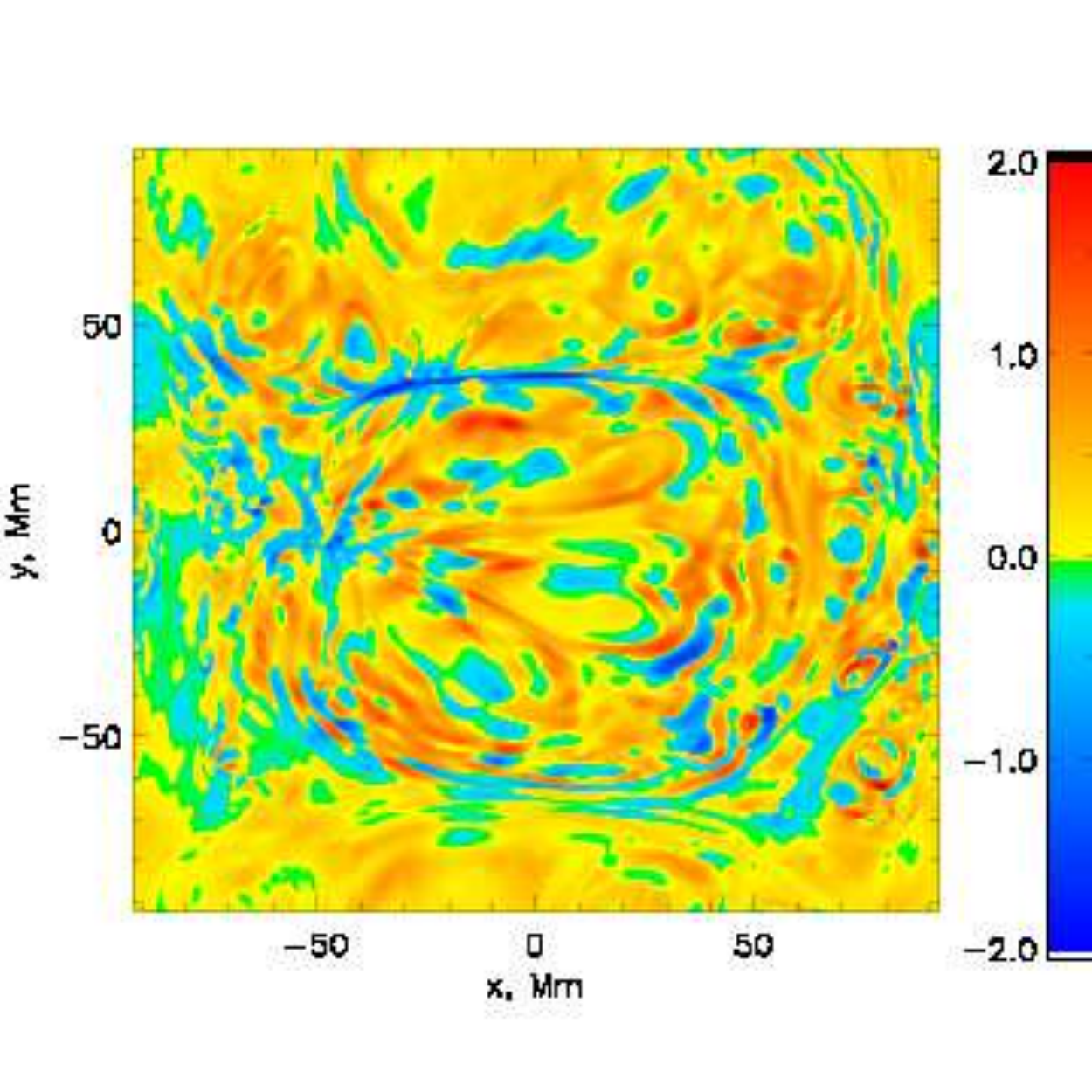}
\vspace{-1cm}

   \includegraphics[width=7cm]{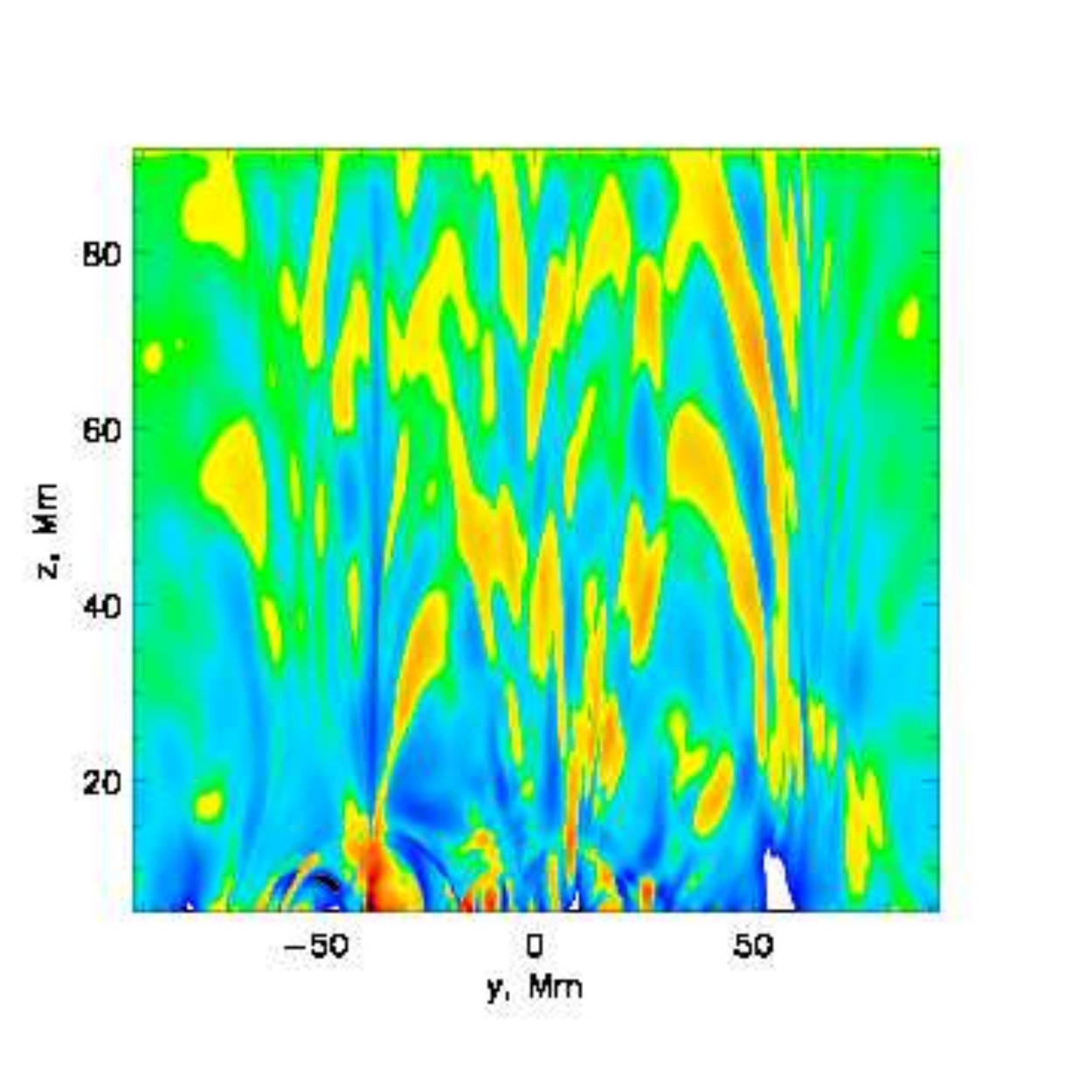}
   \caption{ TOP: Mass flux in a horizontal plane.
             BOTTOM: Mass flux in a vertical plane.
             The data is scaled to $\pm 10^{-7}$ kg m/s.
           }
   \label{massflux.fig}
\end{figure}

The massflux ($m \cdot u_z$) through a horizontal surface has also been investigated. It shows the decrease to be fixed in time, and that the loss rate depends on the height in the atmosphere. The panels in \Fig{massflux.fig} show snapshots of both a vertical and a horizontal plane at t=132.5 min. Averaging the values in the horizontal plane gives a time dependent evolution which shows a near constant depletion of mass with time and a decreasing magnitude as a function of height in the atmosphere.
% values of the average mass loss
%   Exp-1      Exp-2
%  -0.00738  -0.01119
%  -0.00574  -0.00850
%  -0.00461  -0.00670 
The panels in \Fig{massflux.fig} show large variations in the mass flux in space (and time), following a pattern which resamples the previous discussion of the mass density and vertical velocity. Supporting the presence of only wave phenomena in the experiment. 
In the table of Fig.~5 of Paper I the density along a presumed loop is measured. Using the potential magnetic field model it is possible to estimate the magnetic field strength in these region and from this estimate the area change or density change assuming that the same amount of plasma is moving long the field lines all the time. The result of this exercise is shown in \Tab{t3.tab}.
\begin{table}
  \caption[]{The relative change in density from the five boxes seen in Fig.~5 in Paper I. Additionally, the relative change in the density is calculated assuming flux conservation along the field line, using the potential initial magnetic field model.}
  \label{t3.tab}
  \begin{center}
  \begin{tabular}{r c c } %c }
      \hline
      \noalign{\smallskip}
      Index & Fe XII & B-model \\ %& B 901 \\ 
      \noalign{\smallskip}
      \hline
       1 & 1.00 &  1.00 \\           %& 1.00\\
       2 & 0.96 &  0.93 \\           %& 0.73\\
       3 & 0.91 &  0.72 \\           %& 0.57\\
       4 & 0.73 &  0.58 \\           %& 0.47\\
       5 & 0.46 &  0.52 \\ \hline    %& 0.40\\ \hline
   \end{tabular}
\end{center}
\end{table}
This indicates that the observed change in density to first order can be explained by the general expansion of the magnetic field under the assumption that the boxes are represented by nearly the same magnetic field line. 

%________________________________________________________________
\section{Discussion}
\label{dis.sec}
It is important to find the physical mechanism that drives the systematic upflows from the edges of active regions. When this mechanism has been identified we may also have found the mechanism that drives the slow solar wind from the boundary region of coronal holes. For this study we used a numerical 3D MHD model based on an observational magnetic field where HMI data are used as the basis for a potential field extrapolation. To stress the field configuration, a boundary driving derived from a LCT approach is imposed. The results discussed above show that the experiment does not provide a region containing a systematic upflow. This indicates that there is no continuous physical process, in this model scenario, that can drive such a flow, i.e. magnetic reconnection or expanding loop systems. Instead, magneto-acoustic waves are continuously being launched from the lower atmosphere and propagate into the corona. 

The experiment is expected to include most of the important physical conditions for reproducing upflow if these are directly driven by a reconnection process, so why is it failing in doing so? There may be several reasons for this. Here, we discuss the key building blocks and their possible impact on the experiment.

%\begin{figure}
%   \centering
%   \includegraphics[width=6cm]{eps/alpha_0b.eps}
%   \includegraphics[width=6cm]{eps/alpha_1.5.eps}
%   \caption[]{\label{caff.fig} Comparison of the field line structure of the 
%     initial magnetic field. A number of regions have been chosen to indicate 
%     the changes in the magnetic field structure. Left, the potential magnetic 
%     field. Right, the best fitted constant alpha field. The large difference to 
%     the result in \Fig{Fig1.fig} is due to the requirement that the constant alpha model
%     contains flux balance through the base, which here is seen to indicate that
%     the region with the outflow is composed of closed loops. 
%   }
%\end{figure}

%\begin{figure}
%   \centering
%   \includegraphics[width=8cm]{eps/PFSS.eps}
%   \caption[]{\label{PFSS.fig} A PFSS model of the active region. The only open field lines
%       are coloured ?? and shows that the open flux region is located....%
%
%   }
%\end{figure}

% magnetic field model
\subsection{Potential field model} 
The numerical experiment starts from a potential magnetic field model extrapolated from HMI data, while the solar coronal magnetic field is more likely to be closer to a NLFFF with small local alpha values. A NLFFF model requires high quality vector magnetogram observations. Vector magnetogram observations exist for the analysed region, but deriving a NLFFF model was not possible due to poor data quality and a significant imbalance in the magnetic flux at the source surface for this particular region.

Would a NLFFF model be better than a potential model?
First, a NLFFF model contains free magnetic energy, which implies that much less imposed boundary stress may drive the magnetic configuration into a situation where some fraction of the free magnetic energy can be released through a local long duration reconnection process. 
Second, a NLFFF would typically reduce the characteristic height of the closed magnetic field system. A more compact field would provide for a higher Poynting flux input for the same boundary motion, leading to a larger stress of the field and more free energy to be used for driving the upflow.

Would it make a significant difference to the result?
Creating a long duration systematic upflow in a NLFFF system, compared to a potential system, is unlikely to make a significant difference as the magnetic connectivity of the two magnetic fields are related, implying that the boundaries between closed and open field regions can not be significantly different as both extrapolations will contain magnetic null points approximately at the same locations and of the same type. With other words, the topology of the two fields will host comparable connectivity regions for energy accumulation and release.

 Recent numerical simulations of magneto convection \citep{2009LRSP....6....2N} show the photospheric magnetic field to contain much smaller length-scales than at present can be resolved in satellite observations. These smaller scales introduce both shorter characteristic dynamical time- and length-scales of the 3D magnetic field configurations.
This enhanced salt and pepper distribution of the magnetic field on small scales leads to a more complicated local field-line structure. The small scale convection allows for an additional energy input to the lower regions of the atmosphere through flux braiding \citep{1972ApJ...174..499P, 2002ApJ...576..533P} and, therefore, a higher transferee of energy into plasma heating through dissipation of current concentrations than can be inferred from our modelling. How this energy transfers into higher heights on the other hand is not clear, but it may be transferred further up in the atmosphere by MHD-waves that may be dissipated by, for instance, phase mixing \citep{1983A&A...117..220H} providing additional heating of the coronal domain that may be important when including conduction and radiation in the model.

% driver model

\subsection{Driving mode} 
%The imposed driving velocity is critical for the time evolution of the magnetic field and the LCT approach has a number of limitations. 
The LCT approach has a number of limitations, it only measures horizontal velocities based on the tracking of magnetic features. This works fine as long as the magnetic elements are small independent units that move "randomly" around in the 2D plane and the time resolution in the data is sufficiently high to resolve the motions. When the flux concentrations become too large and their structures are too smoothly distributed, the algorithm have problems deriving the correct flow velocities. This is particularly a problem in the active region area, where the derived flow speeds are clearly suppressed compared to the rest of the area.
% It is unlikely that the LCT method can pick up the proper internal flow velocities that are important for the dynamical evolution in the magnetic system. These flows could represent rotational flows that, in special cases, do not change the appearance of the magnetic concentration, but in reality have a significant impact on stressing the magnetic field. 

Due to the kernel size (19 pixels) used for the detection of flows, small scale motions, which seems to be very important for the local evolution of the magnetic fragments when inspecting the HMI data, can not be detected. Ignoring the small scale evolution of the flow field clearly limits the evolution of the magnetic field, and possible also the general energy flux into the corona \citep{2014ApJ...787...87V}. Smaller kernel sizes do provide smaller flow structures, but also introduce a significant increase in the number of locations where excessive flow speeds are derived due to noise/errors in the two compared magnetograms.
 The LCT is, therefore, not representing the real small scale motions in the photosphere that may be important for driving high frequency perturbations into the magnetic configuration.

% As it was stated earlier, the LCT approach seems to be less reliable than it was hoped to be. This is seen by the fact that the flow field in a single pixel is changing, almost randomly, between each of the 45 seconds HMI frames, despite the fact that the smoothing makes the flow field represent structures that, on the sun, will have life times much longer than the time resolution of the observations. This randomness was also seen when trying the balltracking algorithm \citep{2004A&A...424..253P}. This indicates a general characteristics of these types of approach when attempting to obtain flow field information from observational data. 

A physical process missing in the LCT approach, is the emergence and submergence of magnetic flux, which requires a vertical flow velocity. This process continuously takes place on different length scales and is important for the evolution of the coronal magnetic field. Observations of this active region (see Paper I) indicate that no significant flux changes due to large events of emergence or submergence take place around the expected footpoint region of the field lines showing upflows, but this region shows a general decrease in the unsigned flux over time. It can, therefore, not be ruled out that the change in flux is important for driving the upflows and signals that some important effect is taking place somewhere above the photosphere. To get an idea of the change in the boundary structure between the HMI data and the model one can make direct time comparisons. This shows differences in the flux distribution and that this continues to increases with time. The HMI data show a decrease in the unsigned flux with time, while it remains close to constant over time for the numerical experiment. This implies that the two field representations diverge with time and that these differences are most pronounced on small and medium length scales (\Fig{driver_compare.fig}). It, therefore, can not be expected that the simulation result will represent the evolution of the smaller scale magnetic field. As the upflow last for hours to days,  to us, this  indicates that the small scale motion can not be too important for the general driving of these phenomena if we assume it to be driven by a systematic  reconnection process. The proposed driving for the wave motion as mechanism of the upflows also does not seem to directly initiate upflows either.

 The LCT approach has a natural limitation on the velocity structures on scales smaller than the kernel size. This will limit the energy flux into the magnetic field and therefore only provide a lower limit to the local energy flux.

%A more realistic driver may be the {\it DAVE} approach \citep{2010ApJ...715..242F, 2012SoPh..277..153F, 2014ApJ...795...17K}, where changes in the velocity flow are derived using constraints that the imposed flow velocity should as well as possible represent the actual change in the magnetograms. I WOULD SUGGEST TO SKIP THIS BECAUSE THE REFEREE WILL ASK "WHY DID NOT YOU SUE IT THEN?".....
%
%!! something we clearly need to do to be able to make even more realistic
% models in the future.

% atmosphere model
\subsection{Atmospheric model}
The initial atmosphere model is a simple 1D representation, which contains the basic properties of the solar atmosphere. 
%Assuming that the photosphere and corona are located at the same heights for the full horizontal plane is not realistic, as it is known that the local height of the lower atmosphere changes significantly in time and space due to perturbations of the magnetic field from the convective motions. 
Imposing a different hydrostatic model would not have a significant impact on the dynamical result of the experiment. The major difference would be to change the slope of the transition region and by this the wave speed through this region. In terms of creating a systematic upflow, it is unlikely to have a significant impact on the results, as the evolution of the atmosphere is mostly controlled by the structure and stressing of the magnetic field. 

Here, differences in the gas pressure between open and closed magnetic field regions may be important for driving plasma flows when magnetic reconnection between them takes place. This is not, initially, included in the present experiment, but may be build up over time as the model is stressed by the imposed boundary motions. 

% MHD model
\subsection{MHD approach}
The MHD approach used in the calculations is simplified, ignoring optically thin radiation and anisotropic heat conduction in the energy equation, two physical contributions that are important in the solar atmosphere. Including these terms from the start of the experiment would not be simple. Without a direct heating contribution, the upper atmosphere would cool rapidly resulting in an unrealistic coronal model. One could, as others have done \citep{2002ApJ...572L.113G}, drive the initial field configuration to a steady state situation, where the energy input is converted into localised heating throughout the model. At this time the optically thin radiation and anisotropic conduction are activated.  From \citet{2014A&A...564A..12C} it is seen to require on the order of one hour solar time for a closed loop system before a steady state is reached, and possibly much longer for an open field region. This is, therefore, not  a first option in the present case, as we are mainly interested in seeing whether magnetic reconnection can drive the outflow as a response to the the driving.
 If one would include this, the experiment should start at least an hour earlier to allow the field to build up enough complexity to form current structures on various scales. This is needed to provide enough heating through dissipation to balance the effects of heat conduction and optically thin radiation.
 Another problem is the LCT driver, that does not provide a sufficiently correct flow field to represents the actual stressing of the magnetic field. Using only the scalar magnetograms in the process at the same time prevents the method to take into account the continuously ongoing emergence and submergence of flux that with an increasing likelihood takes place in active regions as the investigated time interval increases.

Would this lead to the formation of a systematic upflow of the order of 10--50~\kms\ if the conditions were correct? It is found in some reconnection experiments \citep{2001ApJ...549.1160Y} that energy release in the corona can be channelled down towards the photosphere by heat conduction. This may heat the lower atmosphere resulting in an expansion where the increased plasma pressure creates an upflow creating an atmospheric evaporation scenario. 
This effect was seen in \citet{2014SoPh..289.4501S} where they conducted a 2D simulations where they imposed an energy deposition at the lower boundary, resulting in a slow pressure driven upflow.
If this is the case, then the experiment needs to be updated to include these energy terms and a systematic coronal energy release would be required. The problem with this type of experiments, is the lack of physical explanations that therefore only pushes the understanding to some unexplained mechanism(s).

 The above limitations of including short length scale magnetic field and boundary driving, point in a direction where one assumes that more energy may be transferred into the lower atmosphere. On average this would heat all over the solar surface and is, therefore, unlikely to be responsible for local heating processes that may drive systematic upflows lasting for hours due to an evaporation process. In this context, the energy conduction from the corona to the lower transitions region may locally be important. But this requires a sufficient enhancement in the energy deposition in the corona to allow an additional energy conduction allowing for an evaporation process. In the experiment waves are travelling up through the "open" field region. If inhomogeneities are present in the corona some of this energy may be tapped locally by phase-mixing \citep{1983A&A...117..220H}. This is a more realistic candidate to heat the corona in the open field regions than direct flux braiding \citep{1972ApJ...174..499P}.

%Reconnection point:
\subsection{Location of reconnection}
Where is the reconnection most likely to take place? Having magnetic null points in the model suggests that these are the locations where reconnection can/will take place. But most of these are high in the atmosphere and an ongoing reconnection here would very likely result in a local hot region and would also initiate high velocity outflows. Something like this is seen in \cite{2013ApJ...771...20M} where an emerging flux system interacts with an open field region. The observational characteristics from this type of event clearly does not fit the observations of the slow upflow. The direct reconnection process high in the atmosphere, therefore, seems rather unlikely. This also rules out the pressure driven upflow (atmospheric evaporation) caused as a secondary effect as it is seen in postflare loops. To slow down the process, the reconnection has to take place lower in the atmosphere, where the impact of the plasma heating is much less. Here, the dominating driver of the flows may be the general pressure difference between the two flux regions, where the short closed flux systems typically will have an excess pressure compared to the "open" field regions.

\subsection{Effect of heat conduction}
Including heat conduction in the calculations leads to an energy transfer from the corona to the lower transition region along the magnetic field line structure. In the general situation most of this energy would be radiated away by either optically thin radiation in the hotter regions or by optically thick radiation in the lower atmosphere. On average a balance between the different energy terms exist, but when additional energy is released locally in the corona, it will effectively be conducted down along the field lines. This may lead to an evaporation flow that pushes heated material upwards into the corona. \citet{2014A&A...564A..12C} show this interplay with localised coronal heating on closed field lines leads to an evaporation flow that heats up an emerging loop system.
This scenario may be important for the systematic upflows observed at the edges of active regions. In many observations the upflows typically take place on "open" field lines that, in local models, reaches large heights. An additional localised heating event on these lines are, therefore, required to initiate an evaporation flow. This heating is not easily explained as these field lines contain very little currents as they are typically close to being potential in structure -- a DC mechanism and localised reconnection, therefore, may not seem likely. The alternative is the conversion of wave energy through phase-mixing. This either requires a sufficient density gradient between different field lines to take place within the length scale of the model, or waves with much shorter wavelength than the ones seen in this experiment. 
% The code allows this type of dissipation, but much sharper gradients in the Alfv{\'e}n speed is required to effectively facilitate such a diffusion process. 

%________________________________________________________________
\section{Conclusions}
\label{con.sec}
For the first time a direct data driven model scenario has been used to simulate the dynamic evolution of an active region that was observed to produce a long duration slow upflow event. The model is based on a potential magnetic field extrapolation using HMI data, and it has a complex internal structure defined by the presence of a high number of magnetic null points above the active region. These magnetic nulls would allow for systematic localised reconnection processes that, directly or indirectly, may be able to drive an upflow. The magnetic field is stressed by imposing a photospheric flow velocity based on a LCT of magnetic elements in the HMI data series.  For the conditions in the model, the experiment was not able to reproduce  magnetic reconnection driven systematic upflow found in the observations of NOAA 11123 (see Paper I). Our experiment has a preference for producing wave motions that propagate upwards in the atmosphere with a frequency that needs to be further investigated. However, similar oscillatory motions are seen in a number of observations of active region upflows with typical frequencies on the order of 3--5 min \citep{2010RAA....10.1307G, Ugarte-Urra2011}.

A systematic stressing of the magnetic field by the convective motions can not be ruled out to drive a plasma upflow. Future experiments using a different approach to determine the imposed stressing of the system will be tried. Here especially the ${\it DAVE}$ code \citep{2010ApJ...715..242F, 2012SoPh..277..153F, 2014ApJ...795...17K} approach sounds promising, as its approach will provide a more realistic velocity field, that better reproduces the evolution of the magnetic field.  Also trying to include more terms in the energy equation on longer experiments will show whether this has any influence on the general evolution of the region.

The experiment shows significant wave propagation in the atmosphere directly driven by the imposed boundary driver. As the magnetic setup is complicated, it provides a good opportunity to use a wave analysis approach to see how this can be use as diagnostics for the underlying physical variables. This will be investigated in a future paper.
%
%______________________________________________________________

\begin{acknowledgements}
This work was supported by a research grant (VKR023406) from VILLUM FONDEN.
Brian Walsh for providing the LCT software.
VAPOR softwareImagery produced by VAPOR (www.vapor.ucar.edu), a product of 
the Computational Information Systems Laboratory at the National Center 
for Atmospheric Research. ZH is supported by the China 973 program 
(2012CB825601) and the National Natural Science Foundation of China under 
contract 41404135.
MM is funded by the Leverhulme trust.
Research at Armagh Observatory is grant-aided by the N. Ireland Department 
of Culture, Arts and Leisure and via grants ST/J001082/1 and ST/M000834/1 
from the UK STFC.
\end{acknowledgements}

%-------------------------------------------------------------------

\bibliographystyle{aa}
\bibliography{aa_astroph}

\begin{appendix}

\section{Online material}

\begin{figure}
   \centering
   \includegraphics[width=8cm]{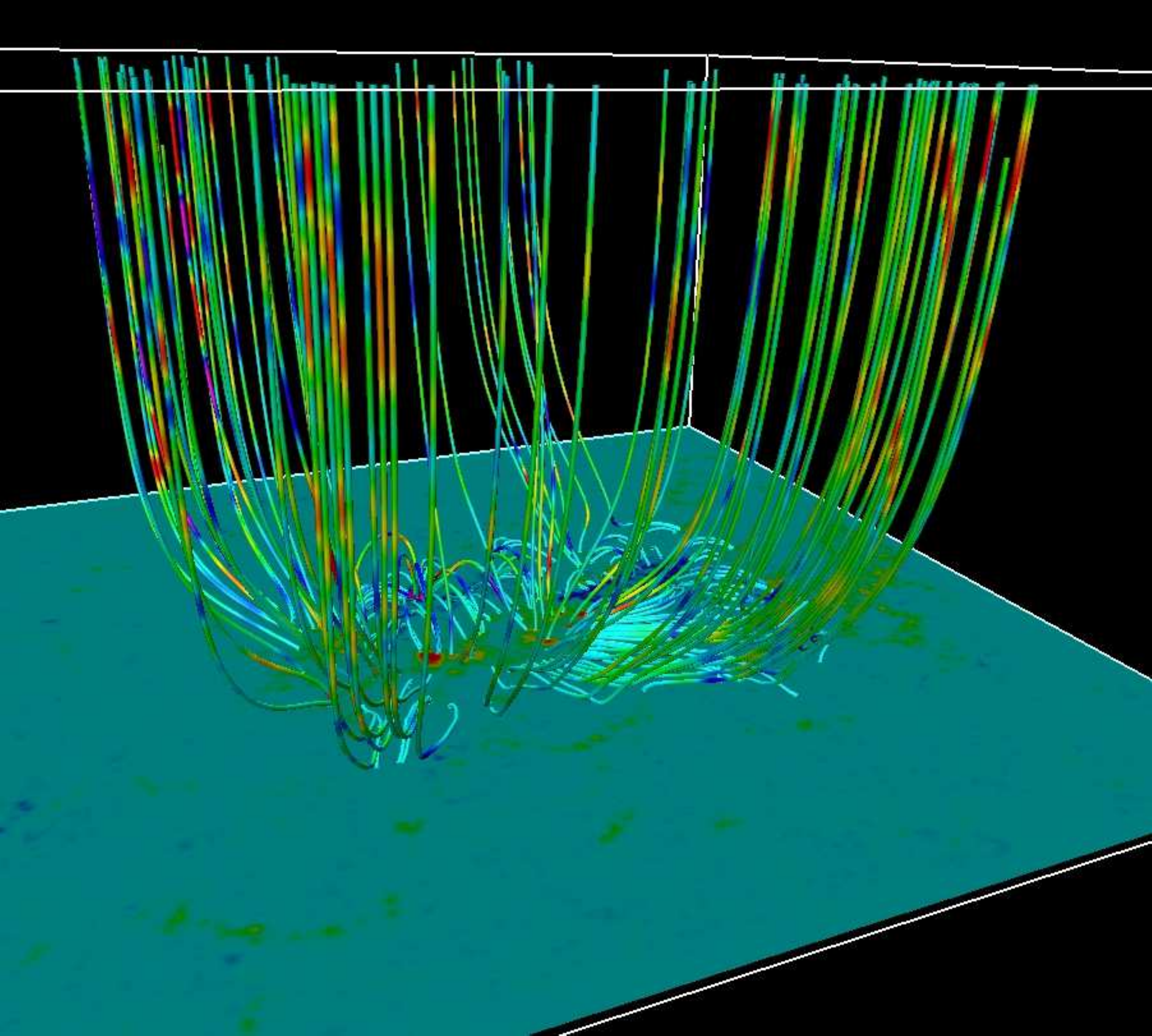}
   \caption[]{The vertical flow-pattern motion upwards along magnetic field lines as a 
              function of time. The base represents the magnetogram, the field lines are 
              traced from a small pillow box around the active region in the photosphere.
              The field lines are coloured with the vertical flow speed, where red is downflow 
              and purple is upflow. The amplitude of the velocity field is scaled to $\pm$ 
              10~\kms. The movie covers the time period from 83.3 to 325 minutes after the start. 
        }
   \label{velocity_field_line.mov}
\end{figure} 

\end{appendix}

\end{document}